\documentclass[11pt]{article}
\pdfoutput=1

\usepackage{jheppub} 

\usepackage[T1]{fontenc} 

\usepackage{tikz}
\usepackage{pbox}

\usepackage{graphicx}

\usepackage{caption}

\usepackage{subcaption}

\usepackage{float}

\usepackage[iso-8859-7]{inputenc}
\usepackage{graphicx}

\def\<{\left\langle}
\def\>{\right\rangle}
\newcommand{\be}{\begin{eqnarray}}
\newcommand{\ee}{\end{eqnarray}}

\def\p{\pi}

\def\vphi{\varphi}

\def\p{\partial}
\def\ls{\left[}
\def\rs{\right]}
\def\lc{\left\{}
\def\rc{\right\}}

\newcommand{\bi}{\begin{itemize}}
\newcommand{\ei}{\end{itemize}}

\title{On the initial conditions for inflation with plateau potentials: the $R+R^2$ (super)gravity case}

\author{Ioannis Dalianis  \, and }
\emailAdd{dalianis@mail.ntua.gr}
\affiliation{Physics Division, National Technical University of Athens, \\  15780 Zografou Campus, Athens, Greece}

\author{Fotis Farakos}
\emailAdd{fotisf@mail.muni.cz}
\affiliation{ Institute for Theoretical Physics, Masaryk University, \\ 611 37 Brno, Czech Republic}

\abstract{
We discuss the initial conditions problem for inflation driven by the vacuum energy of a plateau potential, and in particular the Starobinsky inflation. 
We show that the supergravity embedding of the $R+R^2$ theory naturally decreases the size of the acausal homogeneity, required for the low-scale inflation to occur, thanks to the presence of the dynamical pure supergravitational ``auxiliary'' fields. 
We  examine the evolution of the $R+R^2$ fields within a FLRW Universe. We also find a dependence of the initial conditions problem on the background spatial curvature.
 }

\arxivnumber{1502.01246}

\begin{document}

\maketitle
\flushbottom

\section{Introduction} 

The hot Big Bang model provides a reliable and tested description for the cosmic evolution from at least as early as the time for the synthesis of the light elements, about one second after the Big Bang.  
Nevertheless, problems associated with the adiabatic expansion of the hot early FLRW Universe, such as the entropy, the flatness and the monopole, could not be explained until the postulation of the inflationary cosmology in the beginning of 1980s. \cite{Guth:1980zm}.
Cosmological inflation 
is a phase of accelerated expansion  
assumed to have taken place in the most primal times, once the universe emerged from the Planck epoch, see e.g. \cite{Linde:2005ht, Linde:2007fr}. It  has attracted the interest of the cosmologists for it is a theory with predictive power and can be naturally implemented in microscopic models \cite{Lyth:1998xn}, without the need for rather special initial conditions. Inflation, indeed, is the most efficient mechanism that magnifies, homogenizes and isotropizes the universe and on top of that it successfully describes the CMB temperature fluctuations   according to several data sets; the P{\scshape lanck} 2013 and 2015 results \cite{Ade:2013zuv, Ade:2013uln, Planck:2015xua, Ade:2015lrj} is the latest example. Inflation can be implemented simply by a theory whose matter content acts as vacuum energy,  or equivalently by a quasi cosmological constant term, and magnitude comparable to the GUT scale.

P{\scshape lanck} CMB data \cite{Ade:2013uln, Ade:2015lrj}, although strongly support the basic picture of the inflationary theory may  question  the generality of the cosmological phase because special initial conditions seem to be required \cite{Ijjas:2013vea, Guth:2013sya, Linde:2014nna, Ijjas:2014nta,  Mukhanov:2014uwa}. 
The absence of primordial tensor modes\footnote{Here, we consider that the signal of BICEP2 CMB experiment is contaminated by foreground dust \cite{BICEP2/Keck:2015tva}.} and the spectral index values for the scalar perturbations favour the plateau-like potentials. These potentials are characterized by relatively low energy densities 
 and are not capable to drive an inflationary phase right after the Planck era implying that our Universe started with a decelerating  phase ($\ddot a <0$) instead of an accelerating one ($\ddot a>0$). It is known that low energy scale inflation renders the generality of the initial conditions subject to  speculation,  see e.g. Goldwirth and Piran \cite{Goldwirth:1991rj} for a classical review on this topic.
Motivated by these observations, in this work, we examine the initial conditions required for the $R+R^2$ gravity and supergravity models of inflation.

\subsection{The initial conditions problem}

In the Friedman-Lemaitre-Robertson-Walker (FLRW) model the Universe has a finite age and the cosmological scale factor $a(t)$ grows slower than $t$, facts which lead to the notion of the horizons: the particle and the event horizons that grow linearly with time. Hence, regions in the universe have past histories that do not interact. The present observable universe consists of about $10^5$  causally disconnected regions at the epoch of recombination. However, the CMB is uniform to less than one part in $10^4$.   It has been shown that the class of all initial conditions for which the universe at late times behaves as an FLRW Universe is of measure zero, see e.g. Collins and Hawking \cite{Collins:1972tf}. It is unlikely that the universe began in a chaotic state and has reached the CMB homogeneous state in the course of an adiabatic evolution. This reasoning motivated the introduction and establishment of the inflationary theory.

According to the latest P{\scshape lanck} data\footnote{The discussion and the results of the paper were based on  P{\scshape lanck} 2013 data. This version (v2) also cites the P{\scshape lanck} 2015 data which further support the plateau inflationary potentials.}, the inflationary models fully consistent with the data are the plateau-like potentials with a representative example the Starobinsky model \cite{Starobinsky:1980te}. 
It is an $f(R)=R/2+R^2/(12m^2)$ gravity theory 
which in the dual picture yields a potential for the scalaron $\varphi$ that reads
\begin{equation}\label{starp}
V_{R^2}(\varphi) =V_\text{INF} \left(1- e^{-\sqrt{\frac{2}{3}}\varphi/M_P}\right)^2\,.
\end{equation}
The $V_\text{INF}=(3/4)m^2M^2_P$ is the characteristic upper bound of the inflationary energy density for the Starobinsky model,  $V_\text{INF} \sim 10^{-10}M^4_{P} \ll M^4_{P}$.
This model, although it originally accounts for one of the first attempts to describe the evolution of the universe in its earliest moments, it differs from the dominant pre-P{\scshape lanck} inflationary picture. For decades the standard paradigm for inflation has been that our Universe emerged from the quantum gravity era,  
and  has been magnified to cosmological scales thanks to the prevailing presence of the potential energy of a scalar field in the  energy-momentum tensor. 
The common illustration of this  paradigm has been the quadratic large field model, $V(\phi)=m^2\phi^2$, which is at the edge of the 95\%  CL
contours allowed by P{\scshape lanck}+WP+high-$\ell$ CMB data \cite{Ade:2013uln} (and ruled out at over $99\%$ according to P{\scshape lanck} 2015 data sets \cite{Planck:2015xua}).

\begin{figure} 
\centering
\begin{tabular}{cc}
{(a)} \!\!\!\!\!\!\!\!\! \includegraphics [scale=.7, angle=0]{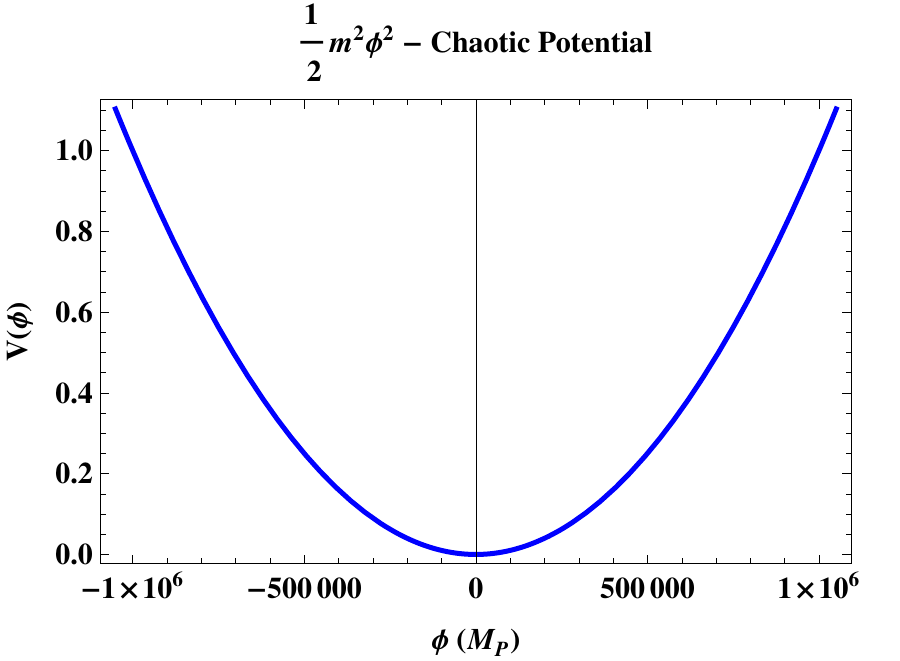} &
{(b)} \!\!\!\!\!\!\!\!\!\!\!\!\!\!\!\!\!\! \includegraphics [scale=.8, angle=0]{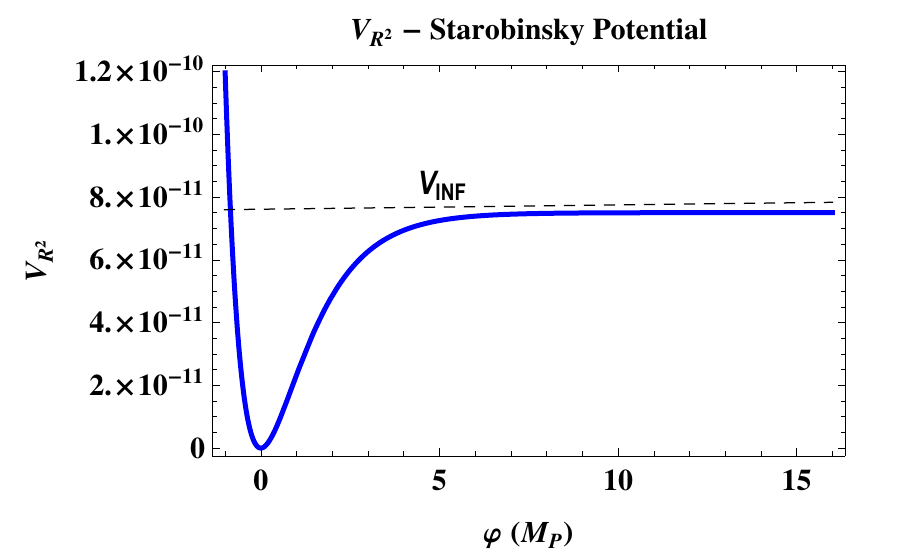}  \\
\end{tabular}
\caption{\small{ In the chaotic models (left panel), inflation starts from Planck densities. In the Starobinsky $R^2$-model (right panel) inflation cannot start at the Planck time because the inflationary sector of the potential is bounded from above, $V_{R^2}<V_\text{INF}\ll M^4_P$. }}
\end{figure}

After P{\scshape lanck}, inflation  appears to start at low energies within a pre-existing homogeneous initial patch. Hence, from one point of view, the plateau potentials bring back the problematic requirement of initial acausal homogeneity a fact that renders inflation nongeneric and reduces 
 some of its appealing power to free cosmology from the need for specific initial conditions.

It is very motivated the inflationary phase  to have been initiated close to the Planck energy scales for it assures the natural creation of our observable Universe without rather special initial conditions.  Indeed, even a fundamentally small initial patch of Planck length radius $ l_P$ 
when dominated by the potential energy of the inflaton field, $\frac12 \dot{\phi}^2 +\frac12(\nabla \phi/a)^2\lesssim V(\phi)\sim \rho_\text{tot}\sim M_P^4$,  starts expanding in an accelerating manner.
The essential implication of this accelerated expansion is the presence of a nearly constant event horizon distance whose size is also  $\sim l_P$, that is, of the order of the curvature scale, the so-called Hubble radius. The curvature scale is the characteristic length and will be also used as {\it unit of length}. For a scale factor dependence $a(t) \propto t^n$ in the time interval $(t,t_{max})$  the event horizon reads for $t\ll t_{max}$
\begin{equation} \label{devent}
d_\text{event}(t, t_\text{max})=a(t)\int^{t_{max}}_t \frac{dt}{a(t)} \simeq 
\begin{cases} 
\frac{n}{1-n} \,H^{-1}(t) \left(\frac{t_{max}}{t}\right)^{1-n}, &  \text{for} \quad n<1\\ 
\frac{n}{n-1}\, H^{-1}(t)\,, &  \text{for} \quad n>1
\end{cases}
\end{equation}
where $H\equiv \dot{a}/a=n/t$ is the Hubble scale. For an accelerated expansion, $n>1$, the event horizon is roughly $d_\text{event}(t) \sim H^{-1}(t)$. Obviously, the dependence is also similar for the $a(t)\propto e^{Ht}$ case. 
The importance of the event horizon is that it {\it protects} the initial smooth patch from the outside inhomogeneous regions where the gradients of the field 
 are nonzero. Otherwise, if the event horizon had been unbounded,  the inhomogeneities would have propagated and infested the initial smooth patch, rendering it unable to accommodate the inflationary phase. 
 
It has been actually shown that homogeneity on super-Hubble scales is required in order for inflation to start \cite{Goldwirth:1991rj, Kung:1989xz, Vachaspati:1998dy}.
Indeed the inhomogeneity due to the gradients of the fields are critical and can prevent inflation. Inflation starts when $V(\varphi)> \frac12(\nabla \varphi/a)^2 \sim (\delta \varphi/a L)^2$ where $L$ the comoving wavelength of the homogeneity and $\delta \varphi$ the typical change in $\varphi$. At the onset of inflation it is $\sqrt{3} H M_P  \sim \sqrt{V}> \delta \varphi/a L$ hence
\be \label{homog-limit}
\frac{aL}{H^{-1}} >\frac{\delta\varphi}{3M_P}\,.
\ee
This qualitative estimation demonstrates that field variations $\delta\varphi \gtrsim  M_P$  cannot have a wavelength smaller than the Hubble radius. Any such inhomogeneity has to have a wavelength $L_\text{ph}=aL> \text{few} \times H^{-1}\equiv \xi \,H^{-1} $.  For inflation to occur the homogeneity has to be assumed on super-Hubble scales \cite{Goldwirth:1991rj, Kung:1989xz, Vachaspati:1998dy}.

\begin{figure} 
\centering
\includegraphics [scale=.985, angle=0]{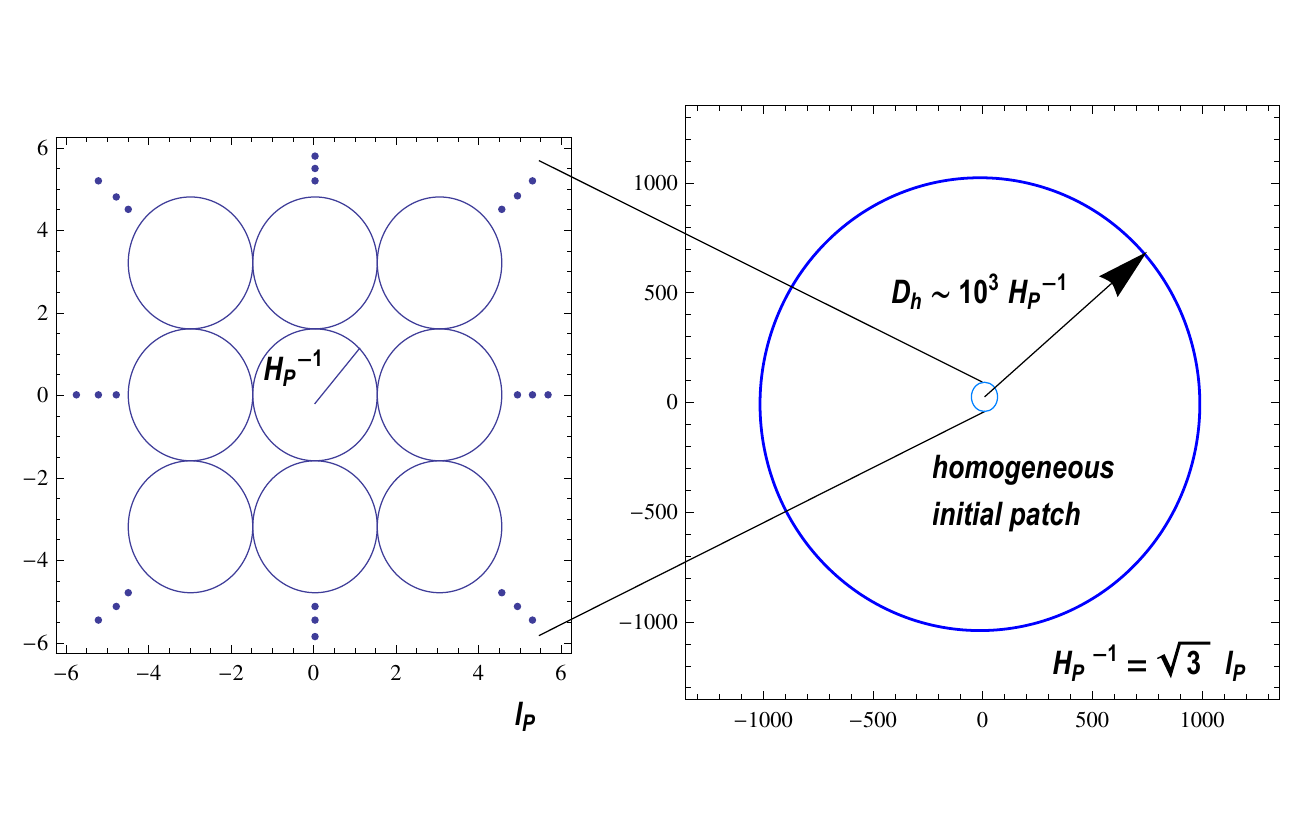} 
\caption{\small{The figure illustrates schematically the initial conditions problem for the plateau inflationary models as the Starobinsky $R^2$ where $V_\text{INF} \sim 10^{-10}M^4_
P$. The delayed inflationary dynamics imply that right after the Planck time hundreds of billions of causally disconnected regions (CDR) have to be homogeneous in order not to spoil the onset of inflation ($D_h\equiv D_\text{homog}(t_P)$).}}
\end{figure}

If inflation is unable to start at energies close to the Planck scale, as it happens at the Starobinsky $R^2$ model, then the minimum size of the initial homogeneous patch has to be much larger than $l_P$. The Starobinsky potential has two regions, a bounded from above plateau region ($\vphi\gg 1$) that drives inflation at energy densities $V_\text{INF} \ll M^4_P$, and a divergent part ($\vphi<0$) which, however, cannot constrain the kinetic energy of the field and no inflation takes place.
Given the smallness of the Starobinsky inflationary plateau energy density, $V_\text{INF}$, one  has to assume that a kinetic-energy domination regime preceded the inflationary phase. In such a case,  $V(\vphi)\ll \frac12 \dot{\vphi}^2 \sim \rho_\text{tot}$, the scale factor grows like $t^{1/3}$ until the domination of the plateau potential yielding an event horizon of size $d_\text{event}(t_\text{init} \sim t_{P}, t_\text{max}> t_\text{INF}) \sim 3\times 10^3 H^{-1}_P$, where $H^{-1}_P \equiv \sqrt{3}\, l_P$ is the Hubble radius at the Planck time. 
Hence, one has to expel the density inhomogeneities at least $10^3$ Hubble scales farther if the Universe has emerged from the Planck densities, $M^4_P \equiv (2.4 \times 10^{18} \text{GeV})^4$. In particular the minimum required homogeneous region is 
\be \label{homcon1}
D_\text{homog}(t_\text{init}) = d_\text{event} (t_\text{init}, t_\text{INF}) + \left[ d_\text{event} (t_\text{INF}, t_\text{max}) +  H^{-1}(t_\text{INF}) \right] \frac{a(t_\text{init})}{a(t_{INF})}\,
\ee
or
\be \label{homcon}
D_\text{homog}(t_\text{init}) = d_\text{event} (t_\text{init}, t_\text{max}) +  H^{-1}(t_\text{INF})  \frac{a(t_\text{init})}{a(t_{INF})}\,,
\ee
where $t_\text{max} \gg t_\text{INF}$ a time deep inside the inflationary era. That is, inflation requires at $t_\text{INF}$ a homogeneous patch of minimum radius 
\be
\xi \, H^{-1}(t_\text{INF}) \, =\, d_\text{event} (t_\text{INF}, t_\text{max}) +  H^{-1}(t_\text{INF})
\ee
which can exist only if the primary patch at $t_\text{init}$, $\xi \, H^{-1}(t_\text{INF})\,a(t_\text{init})/a(t_{INF})$, is surrounded by a {\it supplementary} homogeneous shell of width equal to the event horizon distance $d_\text{event} (t_\text{init}, t_\text{INF})$,  see Fig 3.  The parameter $\xi$ is  of the order $  {\cal O}(1)$. In particular in \cite{Goldwirth:1991rj} the $\xi$ is evaluated to be $\xi > 3/2$ for exponential inflation, i.e with equation of state $w=-1$. 
Here we find $\xi\sim 2 $ for inflation to start at low energies, where $t_\text{INF}$ is the time that the equation of state of the fields drops below $w=-1/3$. 

For flat space and $t_\text{init}\sim t_P$ and $t_\text{INF} \sim V^{1/4}_\text{INF}$ the corresponding initially homogeneous volume, $(4/3)\pi\, D^3_\text{homog}(t_P)$, is at least $10^{11}$ times bigger than $(4/3)\pi l_P^{-3}$ which means that, initially,  hundreds of billions of causally disconnected regions were much similar without any dynamical reason. Briefly we call them  {\it Causally Disconnected Regions (CDR)}. We consider the $l_P$ as the {\it causal horizon} at Planck times. For an open Universe the number of CDR required to be homogeneous is even larger while for a closed Universe the number is decreased about an order of magnitude, albeit the CDR remains formidable large. In fact, these are much special initial conditions for the $R^2$ model and any similar plateau potential inflationary models. 

On the same footing with the CMB homogeneity reasoning,
one concludes,  for a trivial topology,  that it is respectively unlikely that a Universe began in a chaotic state  and has reached the homogeneous state required for the plateau inflation to start.

\begin{figure} 
\centering
\includegraphics [scale=.985, angle=0]{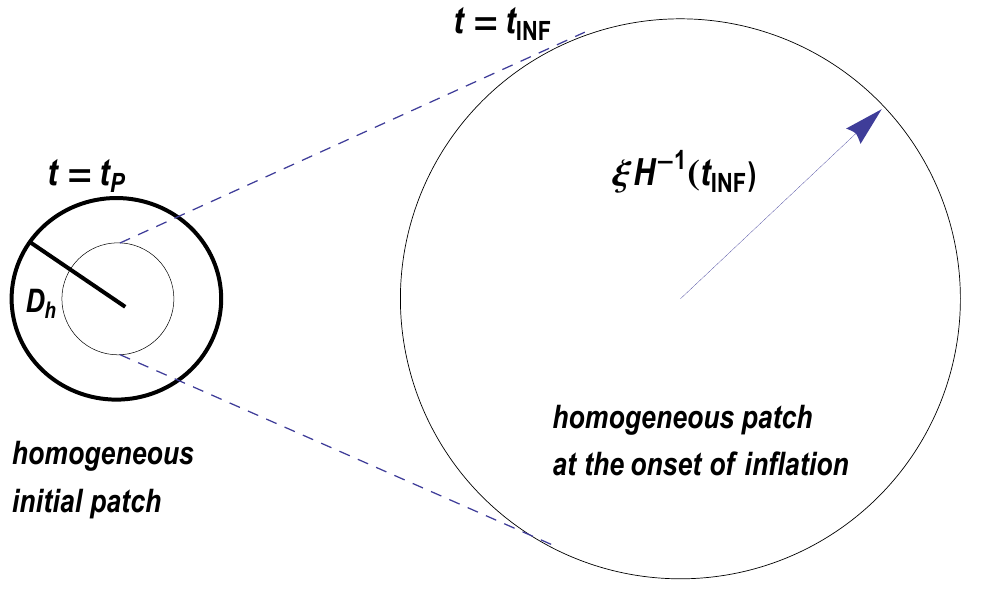} 
\caption{\small{A schematic illustration of the minimum homogeneous patch at two different times: right after the Planck time, $t=t_P$, and at the onset of inflation, $t_\text{INF} \sim 10^{5} t_P$. Its radius at $t_P$ is $D_h\equiv D_\text{homog}(t_P)$ and at $t_\text{INF}$ is $\xi H^{-1}_\text{INF}\sim 10^5 l_P$.}}
\end{figure}

\subsection{Outline of the objectives and results} 

One of the radical implications of the CMB data interpretation is that the $R+R^2$ successfully fit P{\scshape lanck} results \cite{Ade:2013uln} hence, it may provide insight into the effective description  of the fundamental theory of gravity at the particular energy scales. 
In this paper we focus on this possibility examining the $R+R^2$ pure gravity and pure supergravity \cite{Wess:1992cp, Buchbinder:1998qv, Freedman:2012zz}, 
and their status related to the {\it initial conditions problem}. 
Indeed, supergravity, as the low energy limit of string theory, is a well-motivated effective description of gravitation for energies relevant to inflation. Our calculations are performed in the {\it Einstein frame}.

It would be rather straightforward to invoke couplings  to matter and build by hand scalar potentials which exhibit initially a chaotic inflationary phase  (or inflation of the old type) and subsequently the final plateau inflationary phase; or to invoke non-minimal couplings that yield an effective potential which is flat enough to fit the data with inflation starting from Planck densities \cite{Germani:2010gm,Dalianis:2014nwa, Dalianis:2014sqa}. 
However, by turning to such models one abandons the simplicity and universality of pure gravitational (or supergravitational) microscopic description of inflation. In this work,  we focus exclusively on the {\it pure} $R+R^2$ (super)gravity models. 

Our goal is to revisit the problem of the initial conditions 
in a supergravity setup, 
and in the case that the spatial geometry of the primal universe has not been flat. 
Despite the observed flatness of the present Universe we should expect that the initial patch had a non-flat geometry {\it before} the onset of inflation. It is inflation itself that justifies the post-inflationary flatness. We consider open and closed FLRW spatial background geometries.  Other suggestions for non-trivial pre-inflationary topologies can also be found in the literature, see e.g. \cite{Starkman:1998ed, Linde:2004nz} for compact flat or open Universe. 

Minimal supergravity has two different formulations: the old-minimal \cite{Ferrara:1978em,Stelle:1978ye} and the new-minimal \cite{Sohnius:1981tp}. 
One of the objectives is to examine whether the embedding of the Starobinsky model in minimal supergravity renders it more motivated in terms of the assumptions usually requested for the  inflationary initial conditions. We report an affirmative answer to this question: the initial conditions are {\it significantly relaxed, however, not fully addressed}. The reason is the presence of the dynamical pure supergravitational ``auxiliary''\footnote{Actually, the adjective ``auxiliary'' is literally wrong. In $R+R^2$ supergravity these fields propagate and are not auxiliary at all. However, we keep this term using  quotation in order to keep the standard  supersymmetry terminology. } fields.

Let us explain the {\it origin} of these fields. 
In supersymmetric theories there exists a class of bosonic or fermionic fields which do not propagate and are in principle integrated out. Their existence is essential for the off-shell closure of the algebra and for the construction of generic couplings. 
In fact superspace methods by construction give rise to these fields. 
Supergravity theories, being supersymmetric, also include auxiliary fields, 
and the difference in the auxiliary field sector  is  the root of the difference between the old-minimal and the new-minimal formulations. 
When higher curvature terms are introduced, the auxiliary fields might pick up kinematic terms. 
In fact, for $R+R^2$ supergravity this is exactly what happens. 
Therefore, when building $R+R^2$ theories of old-minimal or new-minimal supergravity, 
by construction one will find additional propagating degrees of freedom; 
these are a by-product of supersymmetry and impossible to avoid. 
In this work, instead of treating these fields as merely an eccentricity 
of supersymmetry,  we show that they have important physical consequences. 
Moreover, since we are working with supersymmetric theories, these fields have to reside inside appropriate supermultiplets. 
Indeed, these supermultiplets can be uncovered by turning to the dual description of $R+R^2$ supergravity, 
which is standard supergravity coupled to additional matter fields: 2 chiral multiplets for the case of old-minimal supergravity  \cite{Ferrara:1978rk,Cecotti:1987sa,Kallosh:2013lkr,Farakos:2013cqa,Ferrara:2013wka,Dalianis:2014aya} or one massive vector multiplet for the case of the new-minimal \cite{Cecotti:1987qe,Farakos:2013cqa,Ferrara:2013rsa}. 
To perform the analysis in the following sections we will employ the dual description of the $R+R^2$ supergravity theories, 
but one should bear in mind that these additional superfields have a pure supergravity origin and therefore no additional sector is invoked other than pure supergravity.

Turning to the {\it dynamics} of these fields, although they cannot initially drive inflation, they do implement a relatively fast expansion rate.  Inflation starts naturally after a period of ``kinetic-potential energy balance'' generically yielding much more than 60 e-foldings. The relaxation of the initial conditions is greater in the old-minimal embedding where the dynamics of the ``auxiliary'' fields is described by scalar bosons.  In the new-minimal, the extra bosonic fields include a gauge field that leads to an anisotropic expansion which, though it does not prevent inflation from starting, it ameliorates the initial conditions problem to a smaller extent.
 We also note that although the Universe can start with a potential energy $V(\vphi)$ not too much smaller than the Planck energy density there may be no eternal inflation in the supergravity Starobinsky model.

 A homogeneous FRLW patch features either an open or a closed geometry. For the later case, the conditions for initial homogeneity are translated to conditions for the initial size of the post-Planckian closed Universe because the curvature term has to be always subdominant in order the collapse to be avoided. For the former case, however, the curvature term can dominate. In such a case the 
 radius of  initial homogeneous region though remarkably small the number of the CDR is found to be large again.  
\\
\\ 
In the next section we present the pre-inflationary dynamics of the $R+R^2$ theory embedded in the old-minimal supergravity framework and specify the initial conditions for inflation to start. In section 3 we repeat the analysis for the new-minimal supergavity case where the pre-inflation expansion is anisotropic. Closed and open background FLRW geometries are considered in section 4 and last, in section 5, we conclude.

\section{Old-minimal $R+R^2$ supergravity: the  $V_{\text{sugra}R^2}$ }

\begin{figure} 
\centering
\includegraphics [scale=.595, angle=0]{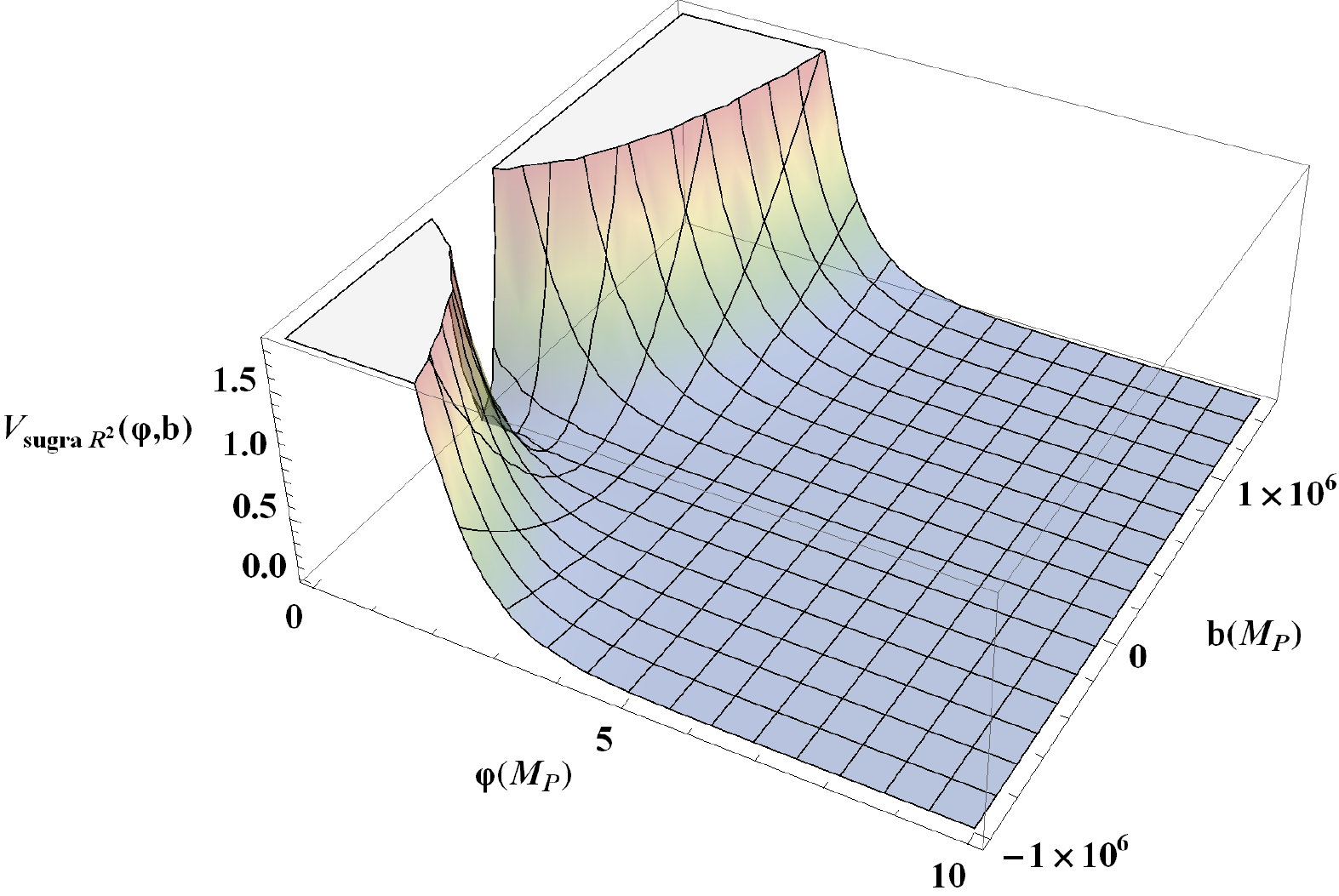} 
\caption{\small{ The shape of the scalar potential for the old-minimal  $R+R^2$ supergravity Starobinsky model. The one direction in field space corresponds to the inflaton $\varphi$ and the second to the $b$-field that makes Planck densities accessible. The potential is depicted in Planck units, $M_P=1$. }}
\end{figure}

The Planck data \cite{Ade:2013uln} favour inflationary models which predict a small amount 
of gravitational waves  (small $r$) and spectral index $n_s=0.96$.
This has given rise to the concern  \cite{Ijjas:2013vea} that inflation is not an invulnerable candidate to solve 
the fine tuning problems of the Big Bang since it may have a fine tuning problem of its own. 
This is related to the fact that models which predict small $r$, 
for example the Starobinsky model of inflation \cite{Starobinsky:1980te} (see \cite{Kehagias:2013mya} for different models with similar effective description during inflation)
have a flat potential
$ V_{R^2}$ that predicts a tensor-to-scalar ratio $r_*=0.0033$ at the pivot scale. For this value of $r$, the energy density of the plateau can be found from the relation $V_*=1.5\pi^2 A_s r_* M^4_P \simeq (0.8 \times 10^{16})^4$ GeV$^4$ $ \simeq 1.2 \times 10^{-10} M^4_P$. This gives rise to a cut-off much smaller than the Planck mass
\be
\label{11}
V_\text{INF} \lesssim 1.2 \times 10^{-10} M_P^4 . 
\ee 
The natural way to overcome this is by somehow being able to have
$V \sim M_P^4$
at the initial stage of the Universe. 
It is easy to see that in the Starobinsky model (\ref{starp}) 
Planck scale potential energy densities can never be realized. 
This leads to a required fine tuning of the inflationary initial conditions. A different approach has been suggested in \cite{Gorbunov:2014ewa}.  
In the following we will show that supergravity offers a natural relaxation to this problem.

The old-minimal supergravity multiplet \cite{Ferrara:1978em,Stelle:1978ye} contains the graviton ($e_m^a$), the gravitino ($\psi_m^\alpha$), 
and a pair of auxiliary fields: the complex scalar $M$ and the real vector $b_m$. 
As we have explained, supersymmetric Lagrangians with curvature higher derivatives also introduce 
kinematic terms for the ``auxiliary'' fields $M$ and $b_m$.   
The embedding of the Starobinsky model of inflation in old-minimal supergravity 
in a superspace (see for example \cite{Wess:1992cp}) approach 
consists of reproducing the Lagrangian 
\be
e^{-1} {\cal L} = -\frac{M_P^2}{2} R+ \frac{M_P^2}{12 m^2} R^2   . 
\ee
This is achieved by \cite{Cecotti:1987sa,Kallosh:2013lkr,Farakos:2013cqa,Ferrara:2013wka,Dalianis:2014aya} 
\be
\label{OM}
{\cal L} = -3 M_P^2 \int d^4 \theta \,  E \,   
\ls 1 -  \frac{4}{m^2} {\cal R} \bar {\cal R}+  \frac{\zeta}{3 m^4} {\cal R}^2 \bar {\cal R}^2  \rs . 
\ee 
Modifications and further properties can be found in  
\cite{Ellis:2013xoa,Ellis:2013nxa,Ellis:2014gxa,Turzynski:2014tza, Kamada:2014gma,Ketov:2014qha, Ferrara:2014yna, Ketov:2014hya,Terada:2014uia,
Alexandre:2013nqa,Alexandre:2014lla}. 
Note  that $R+R^2$ non-minimal (20/20) supergravity  has been constructed in
the linearized level  \cite{Farakos:2015hfa},
using a formalism
which can also describe higher superspin theories \cite{GK,GK2,GK3}.
Lagrangian (\ref{OM}) when expanded to components yields $R^2$ terms and kinematic terms for $M$ and $b_m$. 
One may work directly with (\ref{OM}) but it is more convenient to turn to the dual description in terms of 
two chiral superfields:  ${\cal T}$ and ${\cal S}$. 
In other words, the superspace Lagrangian (\ref{OM}) has a classically equivalent description as 
standard supergravity coupled to additional superfields \cite{Cecotti:1987sa}. 
The equivalent description of the above higher curvature supergravity reads 
\be
\label{standardOM}
{\cal L} =   \int d^2 \Theta \,  2 {\cal E}  \ls \frac{3 M_P^2}{8}  (\bar {\cal D}^2 - 8 {\cal R} )  
e^{- \frac{ {\cal K}}{3 M_P^2}}  \rs + c.c.
+ \int d^2 \Theta \,  2 {\cal E} \,  {\cal W}  + c.c. , 
\ee
with K\"ahler potential
\be
\label{K1}
{\cal K} = - 3 M_P^2 \, \text{ln} \lc 1 +  \frac{{\cal T}  + \bar {\cal T}}{ M_P}  
- 4  \frac{{\cal S} \bar {\cal S}}{M_P^2} 
+ \frac13 \zeta \, \frac{{\cal S}^2 \bar {\cal S}^2}{M_P^4}  \rc  ,
\ee
and superpotential 
\be
\label{W1}
{\cal W} = 6 m \, {\cal T} {\cal S} . 
\ee 
During inflation the universe undergoes a quasi de Sitter phase which implies that supersymmetry is broken. 
In principle the  identification the goldstino supermultiplet even though it is plausible is not always straightforward. 
An inspection of the properties of the model (\ref{standardOM}) during inflation 
shows that the goldstino multiplet is in fact the multiplet ${\cal S}$. 
Moreover since the mass of the sgoldstino becomes large 
it can be integrated out \cite{Lindstrom:1979kq,Farakos:2013ih}, 
leading to a non-linear realization of supersymmetry during inflation as was proposed in  \cite{Antoniadis:2014oya}. 
This amounts to setting \cite{Antoniadis:2014oya} ${\cal S} = X_{NL}$ 
where $X_{NL}^2=0$. 
Of course  this effective description breaks down at the end of inflation. 
This idea has been further developed in \cite{Ferrara:2014kva,Dall'Agata:2014oka}. 
Note that the imaginary component of $T$ is not integrated out due to the non-linear realization.  
It is strongly stabilized during the Starobinsky inflationary phase and therefore does not interfere with the dynamics.

Eventually one finds the effective model
\be
\label{eff}
e^{-1} {\cal L} =  - \frac{M_P^2}{2} R 
- \frac12 \p \varphi \p \varphi  - \frac34 m^2 M_P^2 \left( 1  -  e^{- \sqrt \frac23 \varphi/M_P}  \right)^2   ,
\ee
which is illustrated in Fig. 1. 
We remind that in order to find the Lagrangian (\ref{eff}) from (\ref{standardOM}) one has to set  $S=0$ and Im$T=0$; for these values the fields are strongly stabilized \cite{Kallosh:2013lkr,Farakos:2013cqa,Dalianis:2014aya}.  
From the Planck data \cite{Ade:2013uln} we get
\be \label{m}
m \simeq 1.3 \times 10^{-5} M_P. 
\ee
Now we want the potential energy to become 
${ V} \sim M_P^4  $
for appropriate field values. 
The potential in (\ref{eff}) is unable to do this as we explained. 
\\
\\
In recent work \cite{Dalianis:2014aya} there has been found a new class of {\itshape R-symmetry violating $R+R^2$ models}
which can both provide an inflationary sector and a hidden supersymmetry breaking sector, 
without invoking any matter superfields. 
The new properties of these models which distinguish them from the R-symmetric $R+R^2$ old-minimal supergravity 
is that at the end of inflation the $S$ field contribution starts to become important and the field configuration 
is driven towards the supersymmetry breaking vacuum. 
For these models it is also expected that the initial conditions problem is similar 
to the R-symmetric case that we analyse here.

\subsection{Evolution}

\begin{figure} \label{plot4}
\centering
\begin{tabular}{cc} 
{(a)}\!\!\! \includegraphics [scale=.7, angle=0]{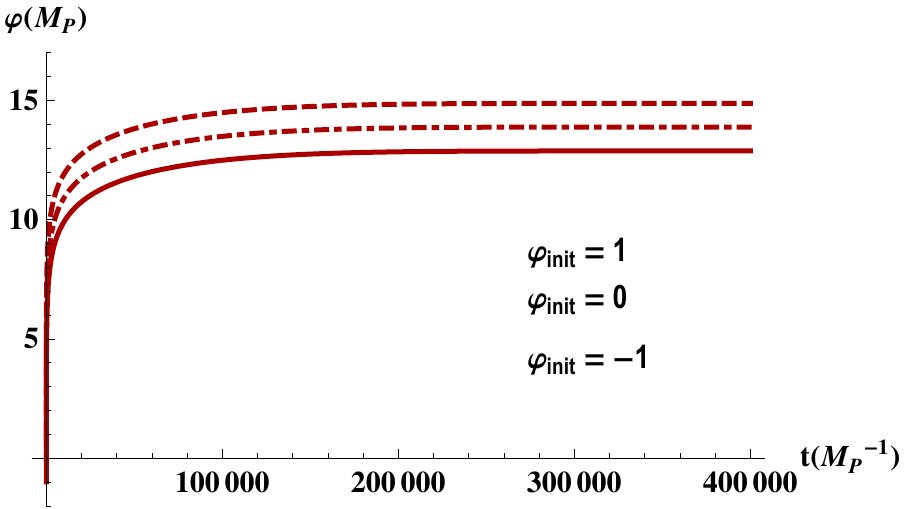} &
{(b)}\!\!\!\!\!\!\!\!\! \includegraphics [scale=.7, angle=0]{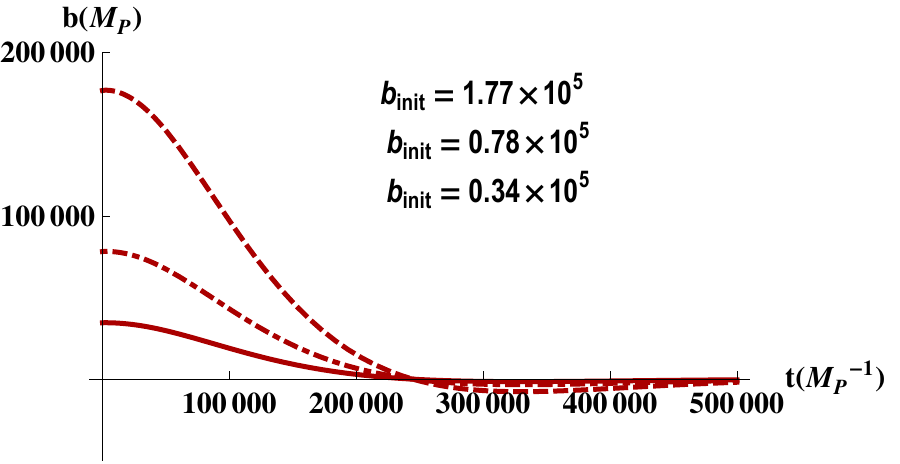}  \\
\end{tabular}
\caption{\small{The evolution of the supergravitational scalar fields $\varphi$ (left panel) and $b$ (right panel) for different initial values and $V(\varphi_\text{init}, b_\text{init})=\rho_\text{kin,init}=0.5 \rho_\text{tot,init}=\,M^4_P$. The first field that evolves is the $\varphi$ while the $b$ remains nearly frozen having an evolution timescale much larger. Once inflation begins around $t\sim 10^5\, t_P$  the $\varphi$-field slowly rolls and the kinetic energy of the $b$-field gets diluted. Initial values for the $\varphi$ about zero can yield many more that $60$-efoldings with $\varphi_\text{INF}>10\, M_P$ and reach the minimum $(\vphi, b)=(0,0)$  after about $10^{10}\,t_P$. }}
\end{figure}

Now we allow for the Im$T$ to take large values. 
First, 
for large Im$T$ values this component also becomes dynamical and after the redefinition 
\be
\begin{split}
\varphi &= \sqrt{\frac32} \, M_P \, {\rm ln}\ls 1 + 2 \, ({\rm Re}T / M_P) \rs,
\\
b &= \sqrt6 \, {\rm Im}T ,
\end{split}
\ee
we have
\be
\label{eff2}
e^{-1} {\cal L} = & - \frac{M_P^2}{2} R 
- \frac12 \p \varphi \p \varphi - \frac12 e^{-2 \sqrt \frac23 \varphi/M_P}   \p b \p b  - V_{\text{sugra}\, R^2}(\varphi, b),
\ee
where
\be \label{oldsugpot}
V_{\text{sugra} R^2}(\varphi, b)= \frac34 m^2 M_P^2 \left( 1  -  e^{- \sqrt \frac23 \varphi/M_P}  \right)^2  
+ \frac12 m^2  e^{-2 \sqrt \frac23 \varphi/M_P}  b^2. 
\ee
Second, the field $S$ remains strongly stabilized and will not affect the evolution. 
Indeed for the mass of the complex field, $S$, 
we have 
\be
\label{mS}
\begin{split}
m_S^2 = &  \frac{e^{-2 \sqrt{\frac23} \vphi} m^2 }{144}  \times 
\\ 
& \left[ 
2 b^2 
\left( 24 + e^{\sqrt{\frac23} \vphi}  \zeta \right) 
+   3 
\left(
24 
+ e^{\sqrt6 \vphi} \zeta 
- 2 e^{2 \sqrt{\frac23} \vphi } (12 + \zeta) 
+   e^{\sqrt{\frac23} \vphi} (48 +  \zeta ) 
\right)
\right], 
\end{split}
\ee
which implies that for a moderately large value of the $\zeta$-parameter \cite{Kallosh:2013lkr,Dalianis:2014aya} 
and for values  for $\varphi$ and $b$ that give Planck-scale energy densities (\ref{I4}) the formula 
for the $S$-mass gives 
\be
m_S^2 > H^2 . 
\ee 
For the rest of our discussion we always assume that when the system evolves, the $S$ field is heavy (\ref{mS}), stabilized at $S=0$, and we do not take it into account for the system evolution. 
This is a rather typical setup in the old-minimal $R+R^2$ supergravity dynamics \cite{Kallosh:2013lkr,Farakos:2013cqa}.

We now turn to a flat FLRW background and study the evolution of the fields and of the spacetime. For 
\be
T^{mn}=\frac{2}{\sqrt{-g}} \frac{\delta\left(\sqrt{-g}\, {\cal L}_{mat}\right)}{\delta g_{mn}},
\ee
the time-time component of the Einstein equation gives 
\be
\label{E1}
3 H^2 M^2_P\, =\, \frac12 \dot \varphi^2 + \frac12 e^{-2 \sqrt \frac23 \varphi/M_P} \dot b^2 + V_{\text{sugra}R^2}(\varphi, b)\,.
\ee
Extremizing the action we take
\be
\frac{\delta {\cal L}}{\delta \phi} -\frac{1}{\sqrt{-g}} \partial_m \left[\sqrt{-g} \frac{\delta {\cal L}}{\delta (\partial_m \phi)} \right]=0,
\ee
where $\phi=\varphi, b$ and $\sqrt{-g}=a^3$. The equations of motion for the fields $\varphi$ and $b$ read
\be \label{f1}
\ddot \varphi +3 H \dot \varphi + \frac{\partial V_{\text{sugra} R^2}(\varphi, b)}{\partial \varphi} =-\sqrt{\frac23} M^{-1}_P e^{-\sqrt{\frac23}\varphi/M_P} \dot b^2,
\ee
\be
\ddot b + (3 H-2\sqrt{\frac23}\dot \varphi M^{-1}_P )\dot b +  e^{2\sqrt{\frac23}\varphi/M_P}  \frac{\partial V_{\text{sugra} R^2}(\varphi, b)}{\partial b}=0\,,
\ee
and in a more analytic form
\be \label{E2}
\begin{split}
 \ddot \varphi &+ 3 H \dot \varphi + \sqrt \frac32 m^2 M_P e^{- \sqrt \frac23 \varphi/M_P} 
\left( 1  -  e^{- \sqrt \frac23 \varphi/M_P}  \right) 
\\
&-  \sqrt \frac23 e^{-2 \sqrt \frac23 \varphi/M_P} \left( m^2 b^2 - \dot b^2 \right)= \, 0,
\end{split}
\ee
and
\be
\label{E3}
\ddot b + 3 H \dot b -2  \sqrt \frac23 \dot \varphi \dot b + m^2 b= \, 0. 
\ee
We want to study the evolution of this system 
which has been also discussed in \cite{Ferrara:2014ima,Kallosh:2014qta,Hamaguchi:2014mza} and  in a different context in \cite{Lalak:2007vi}.

We choose initial conditions such that the initial energy density, $\rho_\text{init}=M^4_P$, is equally partitioned between the kinetic and the potential terms,
\be  \label{I4}
 V(\varphi_\text{init} , \,  b_\text{init})\, =\,\rho_\text{kin, init}\, = \, \frac12 M_P^4 .  
\ee 
An {\it example} of such a set of values that realize these initial conditions is,
\be
\label{I5}
(\varphi_\text{init}, b_\text{init}) = ( 0,  \frac{M^2_P}{m}) \quad \text{and} \quad  \dot \varphi_\text{init}=  \dot b_\text{init} = \frac{1}{\sqrt{2}}M^2_P\,.
\ee

The equations (\ref{E1}), (\ref{E2}) and (\ref{E3}) can be solved numerically. 
For the initial conditions (\ref{I5}) we have found exact numerical solutions, 
which are illustrated in Fig. 5. 
It is easy to see (also from the potential) that the $\varphi$ field, due to the large $b$  values, 
has locally a run-away potential and starts to grow. Later, the $b$ field starts to roll down its potential. 
This is the  {\it imaginary} Starobinsky phase \cite{Ferrara:2014ima,Kallosh:2014qta,Hamaguchi:2014mza}. 
At some point the $b$ becomes small and then the $\varphi$ field 
stops increasing and we have the inflationary initial values
\be
(\varphi_{\rm INF}, b_{\rm INF})  \sim (13 \, M_P ,0)
\ee
for the (\ref{I5}) initial Planck-scale energy densities. 
Then a Starobinsky inflationary phase starts and is
described by the Lagrangian (\ref{eff}), 
which naturally lasts for much more than 60 e-foldings\footnote{We note that there is no fine tuning problem here similar to that of hybrid inflation where special initial values for the fields are required in order  $N\gtrsim 60$ e-foldings to be achieved \cite{Tetradis:1997kp, Mendes:2000sq, Easther:2013bga, Easther:2014zga}.}.
\begin{figure} 
\textbf{  Evolution of the equation of state  \,\,\,\,\,\,\, Evolution of the scale factor}
\centering
\begin{tabular}{cc}
{(a)} \!\!\!\!\!\! \includegraphics [scale=.6, angle=0]{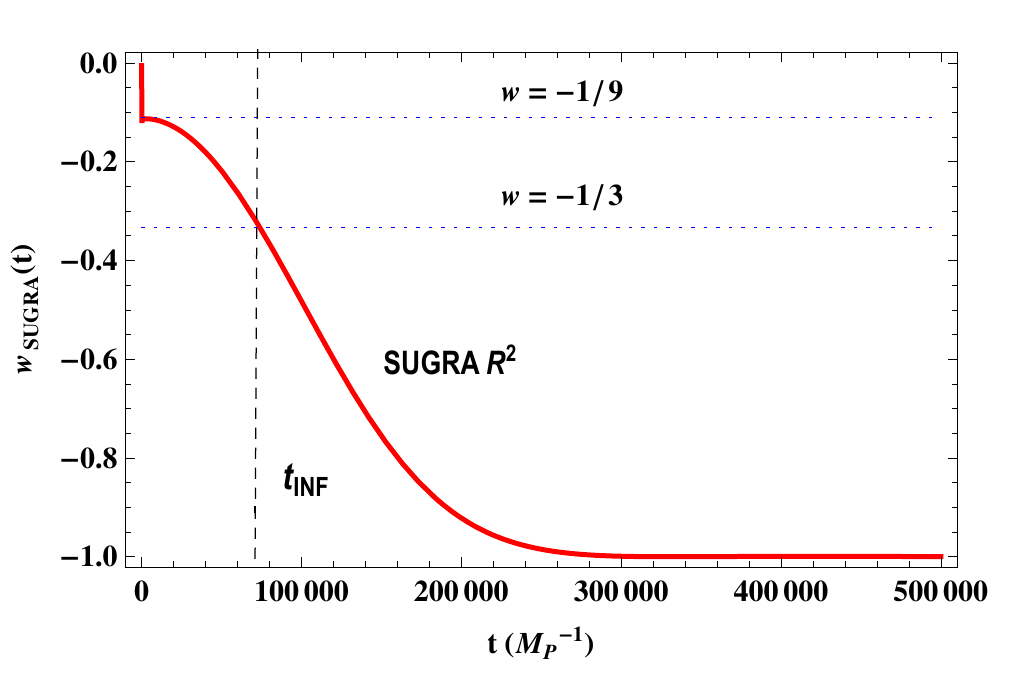} &
{(b)} \!\!\!\!\!\!\!\!\! \includegraphics [scale=.6, angle=0]{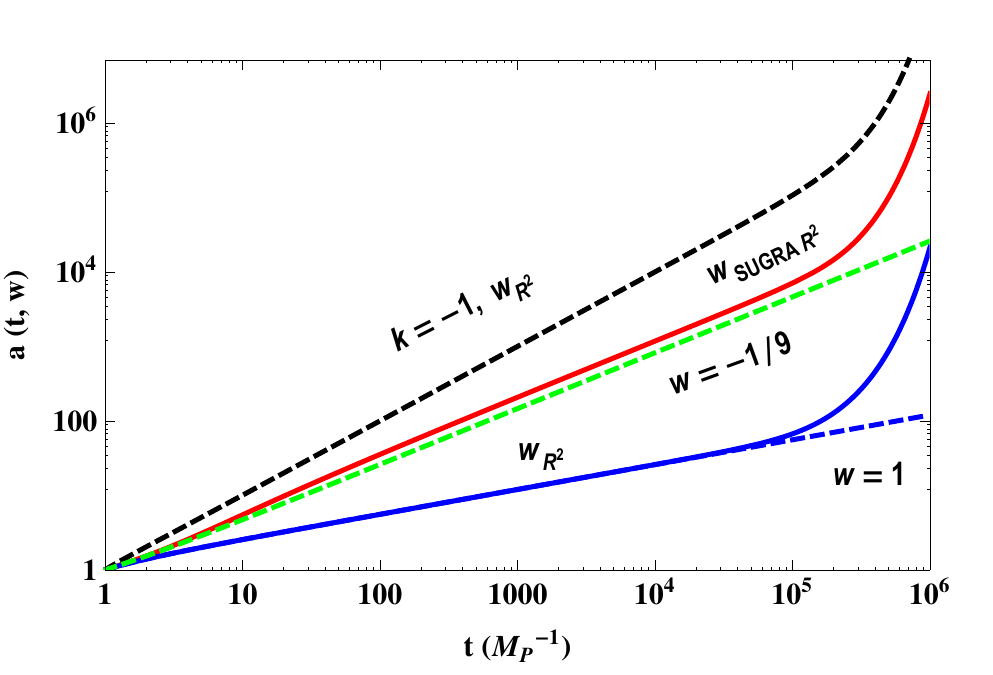}  \\
\end{tabular}
\caption{\small{ {\it The left panel} shows the equation of state, $w$, for the supergravitational system of fields. The initial conditions chosen are equipartition of energy between kinetic and potential thus $w=0$ initially. For some period it is $w\sim -1/9$ and at $t_\text{INF}\simeq 0.7 \times 10^5\, t_P$ the equation of state becomes $w\simeq -1/3$ and acceleration starts. The nearly de-Sitter phase $w\simeq -1 $ starts after $3 \times 10^5 t_P$.  {\it The right panel} shows the evolutions of the cosmological scale factor. The solid lines correspond to solutions for the scale factor of the conventional Starobinsky (lower, blue) and the Starobinsky supergravity (upper, red). The initiation of the accelerating phase is apparent after $t_\text{INF}$. The dashed lines close to the solid ones are the constant equation of state approximations. The lower blue dashed corresponds to the $w=1$ equation of state and describes exactly the evolution of the scale factor before inflation for the $V_{R^2}$ case; the green dashed corresponds to constant $w=-1/9$ which approximates well the $V_{\text{sugra}R^2}$ case until, roughly, the onset of inflation.  The upper black dashed line corresponds to background spatial geometry of negative curvature for the conventional Starobinsky plateau inflationary potential.}}
\end{figure}

To summarize, the dynamical evolution the pre-inflationary stage consists of two phases: 
\begin{enumerate}
\item From  $V_{\text{sugra}R^2} \simeq M_P^4 $ to $ V_{\text{sugra} R^2} \gtrsim m^2 M_P^2 $, 
both $\varphi$ and $b$ participate in the evolution with the $\vphi$ field rolling first.
\item  At  $ V_{\text{sugra} R^2} \simeq m^2 M_P^2 $ starts the standard Starobinsky inflationary phase with $\varphi$ driving inflation, 
and $b$ now strongly stabilized and  integrated out. 
\end{enumerate}

\subsubsection {Dynamics of the expansion} 
The local conservation of the energy-momentum, $\nabla _m T^{mn}=0$, gives the evolution of the energy density which in an FLRW background  reads $\dot \rho+3H(\rho+p)=0$. When the energy density is dominated by a fluid with constant equation of state, $w=p/\rho$, the energy density changes with the cosmological scale factor as $\rho(a)=\rho_\text{init} (a/a_\text{init})^{-3(1+w)}$. From the Friedmann equation we take 
\be \label{scalef}
a(t) =a_\text{init}\left[\frac32(1+w) \sqrt{\frac{\rho_\text{init}}{3\, M^2_P}}(t-t_\text{init})+1 \right]^{2/[\,3(1+w)]},
\ee
where $a(t_\text{init})=a_\text{init}$, $\rho(a_\text{init})=\rho_\text{init}$ for $t_\text{init}\leq t$. When the small constant term in the brackets is negligible, we see that the scale factor changes as $a(t)\propto t^{2/[\,3(1+w)]}$ and the energy density as 
\be \label{rho}
\rho \simeq \frac{4\, M^2_P}{3(1+w)^2} \, t^{-2},\quad\quad \text{for} \quad  t \gg t_\text{init}\,.
\ee
The time $t_\text{INF}$ that signals the onset of inflation is set by the energy density of the plateau. When the total energy density is $\rho \sim V_\text{INF}$ then inflation starts and from 
eq. (\ref{rho}) we take that 
\be
t_\text{INF} \sim  \left[\frac{4\, M^2_P}{3(1+w)^2} V^{-1}_\text{INF} \right]^{1/2} \sim \, 10^{5} M^{-1}_P\, =\,10^{5} t_P\,.
\ee
The precise time is actually smaller because acceleration starts when $w<-1/3$ that is when the kinetic energy density is half the potential energy density, hence $\rho_\text{INF} > V_\text{INF}$ and $t_\text{INF}<10^{5} t_P$. We find numerically that for the  $R+R^2$ supergravity theory inflation starts when 
\be
t_\text{INF} \simeq 0.74 \times 10^{5} M^{-1}_P, 
\ee
for equipartitioned initial energy densities.

The system of the fields $(\varphi, b)$ starts from nonzero values such that  $V(\varphi_\text{init}, b_\text{init})\sim M^4_P$. Due to the small mass of the $b$ field, $m_b\ll H$, the $\varphi$ will roll down the potential (\ref{oldsugpot}) which for constant $b$ has the form
\be \label{appr}
V_{\text{sugra} R^2}(\varphi) \sim V_0
 e^{-2\sqrt{\frac{2}{3}}\varphi/M_P}\,.
\ee
Initially, the approximation $\dot b\sim 0$ is a good one according to the numerical results. For the potential (\ref{appr}) the Friedmann equation (\ref{E1}) and the equation of motion (\ref{f1}) have an exact solution of power law form given by the expressions \cite{Liddle:1988tb}
\be \label{approxsugra}
\begin{split}
a \propto a_\text{init} t^n\,, \quad n=3/4 \\
\varphi=\sqrt{\frac32}\ln\left(\sqrt{\frac{16}{15}\frac{V_0}{M^2_P}} \,t\right)\,.
\end{split}
\ee
According to (\ref{scalef}) we see that $n=2/[3(1+w)]=3/4$. This corresponds to a barotropic fluid with equation of state $w=-1/9$, that is a negative pressure. Numerically we find that in the pre-inflation period the energy density decreases with a slower rate,
\be \label{sugrabef}
w_{\text{sugra} R^2} \lesssim -1/9, \quad a_{\text{sugra}R^2}(t) \gtrsim  t^{3/4}, \quad \rho_{\text{sugra}R^2} \gtrsim \rho_\text{init} a^{-8/3}\,,
\ee
because the actual system is a two-field one.

\subsubsection {Selfreproduction}
At this point we would like to briefly comment on the initial conditions for the $R+R^2$ (super)gravity model that lead to the eternal process of {\it selfreproduction}. After $t_\text{INF}$ the $\varphi$-field rolls pretty slowly, as depicted in the  equation of state $w$ at Fig. 6, and the average quantum fluctuation  $|\delta \vphi| \simeq H/2\pi  \simeq m/(4\pi)$ inside the domain of radius $H^{-1}$ can be larger than the classical variation of the inflaton  $|\Delta \vphi| \simeq \sqrt{3/2}\, M_P/N(\vphi)\,=\, 2\sqrt{2/3}\, M_P \,e^{\sqrt{2/3}\,\vphi/M_P}$ during a Hubble time, where $N(\vphi)$ is the number of e-foldings before the end of inflation. 
Eternal inflation takes place for $|\delta \vphi|> |\Delta \vphi|$ or equivalently for sufficiently large $\vphi_\text{INF}$ values
\begin{equation}
\vphi_\text{INF} > 17.5 \, M_P\, \equiv \vphi^*_\text{INF}\,,
\end{equation}
where the value (\ref{m}) for $m$ has been plugged in. For example, when the initial values of the field is $\phi_\text{init}\sim 0$ and $V_\text{init}\sim 1$ then inflation never enters the self-reproduction regime. 
As it is listed in the table 1, in the old-minimal embedding, only values larger than  $\vphi_\text{init}\gtrsim 13.5$ lead to the self-reproduction regardless the fact that initially the potential energy density is of the order of the Planck scale.

\subsection {Initial conditions}

The standard lore of inflation is that it started right after  
the Planck era and magnified the volume of the Universe $e^{3N}$ times, where  $N\gtrsim 60$ the number of e-foldings.

If inflation begins at times $t_\text{INF} \gg t_P$, i.e. energy densities $V\ll M^4_P$, as it happens with the plateau potentials,  then in the period between  the Planck time and $t_\text{INF}$ the equation of state of matter in the Universe has to be $w>-1/3$. If $w$ is nearly constant then the scale factor increases like $a(t)\propto t^n$ where $n=2/(3+3w) <1$. Then the part of the space that will be in causal contact until inflation begins has radius
\begin{equation}
\begin{split}
 d_\text{event}(t_\text{init}, t_\text{INF}, w) \,&= \, a(t_\text{init})\int^{t_\text{INF}}_{t_\text{init}} \frac{dt}{a(t)} \simeq 
\frac{n}{1-n} H^{-1}(t_\text{init}) \left(\frac{t_\text{INF}}{t_\text{init}} \right)^{1-n},
\end{split}
\end{equation}
for $t_\text{init} \ll t_{INF}$. 

\begin{figure} 
\textbf{\,\,\,\,\quad\quad The event horizons\quad\quad\quad\quad}
\centering
\begin{tabular}{cc}
{(a)}\!\!\!\!\!\! \includegraphics [scale=.65, angle=0]{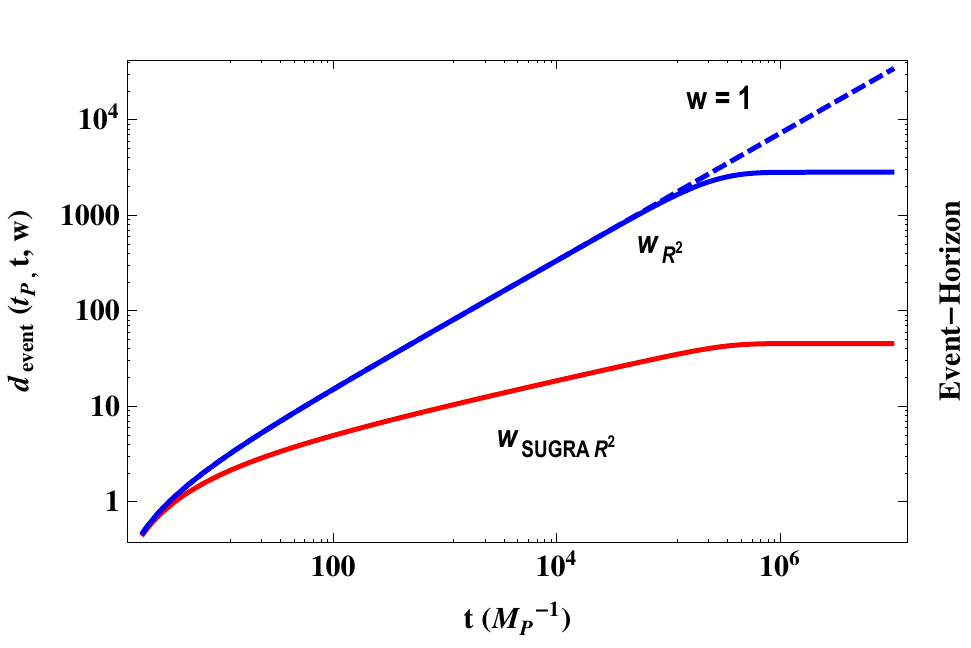} &
{(b)}\!\!\!\!\!\! \includegraphics [scale=.65, angle=0]{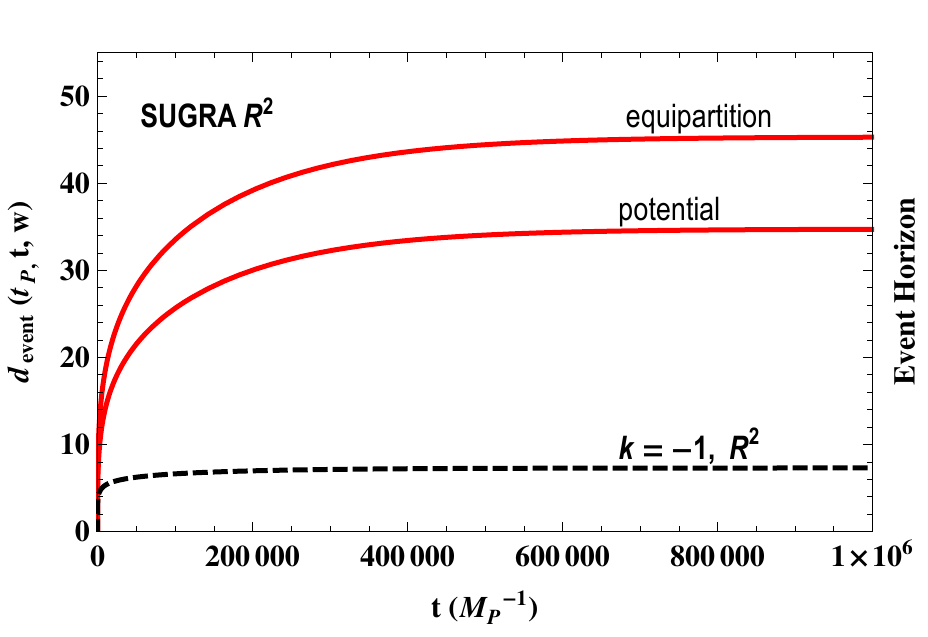}  \\
\end{tabular}
\caption{\small {The figures show the event horizon distances for the Starobinsky (blue) and supergravity Starobinsky (red) model. In the {\it left panel}, in a logarithmic plot, we see that the event horizon for the supergravity is about two orders of magnitude less than the conventional Starobinsky and it becomes nearly constant after $t_{INF}$; this is evident from the dashed line that corresponds to $w=1$ and describes the pre-inflationary evolution of the Starobinsky model. The {\it right panel} shows the event horizon for the supergravity Starobinsky for different initial values for the fields. When the $\rho_\text{tot, init}$  is entirely potential energy then the event horizon distance is smaller than in the equipartition case. The lower black dashed line shows the event horizon distance when the negative spatial curvature dominates. The event horizon distance is in $H^{-1}_P=\sqrt{3} \, l_P$ units.}}
\end{figure}

\subsubsection{$R+R^2$}
In the non-supersymmetric Starobinsky,  the energy density has to be dominated by the kinetic term $\dot{\vphi}^2/2$ as long as  $\rho>V_\text{INF}$ which translates into an equation of state $w=1$. The same is true for all the similar single-field plateau models where $V_\text{INF} \sim 10^{-10} M^4_P$. In such  cases the energy density falls fast as $\rho \propto a^{-6}$ while the expansion is especially small $a(t)\propto t^{1/3}$.  This implies that the initial patch is sensitive to inhomogeneities far away because  the intervening space expands too slowly. In particular the event horizon radius, when we integrate from the Planck time till the onset of inflation, $t_\text{INF} \simeq 0.6\times 10^5\,t_P$, is 
\begin{equation} \label{dkin}
d_\text{event} (t_\text{init}=t_P,\, t_\text{INF},\, w_{R^2}) \simeq \frac12 H^{-1}(t_P) \left(\frac{t_\text{INF}}{t_P} \right)^{2/3} \simeq 766 \, H^{-1}_P 
\end{equation} 
where $H^{-1}_P=\sqrt{3}\, l_P$. 
Actually, the event horizon continues increasing after the initiation of the accelerating phase, $t_\text{INF}$, though much more slowly, and does not get a constant value until the equation of state becomes  $w \cong -1$, see Fig. 6 and 7. We find numerically that the total  event horizon radius at the Planck time is
\be \label{dkin2}
d_\text{event} (t_\text{init}=t_P,\, t_\text{max}, \, w_{R^2}) \simeq 2820 \, H^{-1}_P\simeq 4884 \,l_P\, , 
\ee
where $t_\text{max} \gg t_\text{INF}$ a moment inside the inflationary era.
Hence, the minimum initially {\it homogeneous} region required for inflation to start has radius
\be
\begin{split}
D_\text{homog}(t_P, \, w_{R^2}) 
& \, = \, d_\text{event} (t_P, t_\text{max}) +\, H^{-1}(t_\text{INF})\frac{a(t_P)}{a(t_{INF})}\\
& \sim \, 4884\,l_P + \,H^{-1}(t_\text{INF}) \,\frac{1}{47}  \, \sim \,  8.7 \times 10^3 \, l_P
\end{split}
\ee
where $a(t_\text{INF})\sim 47 \, a(t_P)$ and $H^{-1}(t_\text{INF})\sim 3\,t_\text{INF}$. The minimum number of the causally disconnected regions (CDR) required to be homogeneous is
\be \label{fCDR}
\frac{V_\text{flat}(D_\text{homog}, w_{R^2})}{V_\text{flat}(l_P)} = \frac{\frac43 \pi D^3_\text{homog}}{\frac43 \pi l^3_P} \sim (8.7 \times 10^3)^3 \, \sim \, 7 \times 10^{11} \,\, \text{CDR}\,,
\ee
which manifests the initial condition problem for the plateau potentials such as the Starobinsky $R^2$ model.

\subsubsection{$R+R^2$ supergravity}
On the other hand, in the Starobinsky supergravity model the pre-inflation expansion of space is much faster (\ref{sugrabef}) and the event horizon is much smaller. An approximation (\ref{approxsugra}) is to consider a constant equation of state $w=-1/9$ which yields an event horizon radius $d_\text{event} \propto (t_\text{max}/t_P)^{1/4} \sim 87\, l_P$ for $t_\text{max}=10^5\,t_P$.
An exact result can be obtained numerically for the varying equation of state $w_{\text{sugra}R^2}$ (\ref{sugrabef}). When we integrate from the Planck time until the beginning of inflation, which is found to be $t_\text{INF}=0.74  \times 10^5\,t_P$, we numerically take 
\begin{equation} \label{dpot}
d_\text{event} \left(t_\text{init}=t_P,\, t_\text{INF},\, w_{\text{sugra}R^2}\right)  \simeq 29 \, H^{-1}_P\,. 
\end{equation} 
As in the $R+R^2$ gravity case, the event horizon increases as long as $w>-1$. It remains constant  when the field configuration lies in the plateau with vanishing kinetic energy,  $w \cong -1$, see Fig. 6 and 7. 
The numerical value of the total event horizon reads
\be
d_\text{event} \left(t_\text{init}=t_P,\, t_\text{max}, \, w_{\text{sugra}R^2}\right) \simeq   46 \, H^{-1}_P\simeq 80\,l_P\,.
\ee
The minimum initially {\it homogeneous} region required for inflation to start in the supergravity case has radius
\be
\begin{split}
D_\text{homog}(t_P, \, w_{\text{sugra}R^2}) 
& \, = \, d_\text{event} (t_P, t_\text{max}) + \, H^{-1}(t_\text{INF})\frac{a(t_P)}{a(t_{INF})}\\
& \sim \, 80\,l_P + \, H^{-1}(t_\text{INF}) \,\frac{1}{5407}  \, \sim \, 98 \, l_P
\end{split}
\ee
where $a(t_\text{INF})\sim 5407\, a(t_P)$ and $H^{-1}(t_\text{INF})\sim 4/3 \,t_\text{INF}$. That is, right after the Planck time 
the initial homogeneous volume is required to have radius at least 68 times the Planck length. The minimum number of the CDR is here
\be
\frac{V_\text{flat}(D_\text{homog}, w_{\text{sugra}R^2})}{V_\text{flat}(l_P)} = \frac{\frac43 \pi D^3_\text{homog}}{\frac43 \pi l^3_P} \, \sim (98)^3 \, \sim \,  10^{6} \,\, \text{CDR}\,.
\ee
Compared to the non-supersymmetric case, in the $R+R^2$ supergravity the required initial homogeneous volume is about half a million times smaller,
\begin{equation}
\frac{\#\, \text{CDR}_{R^2}}{\#\, \text{CDR}_{\text{sugra}R^2}}\, = \, \frac{D^3_\text{homog} (t_P,\,  w_{R^2})}{D^3_\text{homog}(t_P,\, w_{\text{sugra}R^2})} 
\sim  \,7 \times 10^5\,.
\end{equation} 

To outline, in the $R+R^2$ supergravity the initial conditions problem though {\it significantly ameliorated} (about one million times) it {\it persists}. 
The evolution of the event horizon for the Starobinsky supergravity as a function of time and the initial conditions (the {\it energy partition} between kinetic and potential) can be seen in Fig. 7.

\section{New-minimal $R+R^2$ supergravity:  the  $V_{\text{nsugra}R^2}$ }

The new-minimal supergravity multiplet \cite{Sohnius:1981tp} contains the graviton field $e_m^a$, 
the gravitino $\psi_m^\alpha$ which are physical fields, a real auxiliary vector $A_m$ which gauges the $U(1)$ R-symmetry and a 
two-form auxiliary field $B_{mn}$. 
The two-form appears here only through the dual of its field strength $H_m$. 
In the theory with no curvature higher derivatives $A_m$ and $H_m$ are integrated out. 
This does not happen when we introduce higher curvature terms. 
In this case a combination of these fields becomes propagating \cite{Cecotti:1987qe}. 
The Starobinsky model of inflation in new-minimal supergravity  reads in superspace  \cite{Cecotti:1987qe,Farakos:2013cqa,Ferrara:2013rsa} 
\be
{\cal L}= - 2 M_P^2 \int d^4 \theta \, E V_{\text{R}} + \frac{\alpha}{4} \int d^2 \theta \, {\cal E} W^2(V_{\text{R}}) + c.c.,
\ee
and in component form for the bosonic sector we find 
\be
\label{star-new}
e^{-1} {\cal L} =  M_P^2 \, \left(- \frac{1}{2} R+ 2{ A}_a H^a-3H_a H^a\right)  
+ \frac{\alpha}{8} \left(-R+6H^2\right)^2  - \frac{\alpha}{4} F^2({ A}^-) ,
\ee
for $ A_m^-={ A}_m-3H_m$. 
It is easy to verify from (\ref{star-new}) that when there are no curvature higher derivatives present, e.g. the limit $\alpha \rightarrow 0$, 
the fields $A_m$ and $H_m$ vanish on-shell. 
The Lagrangian  (\ref{star-new}) is classically equivalent to a Lagrangian where no higher derivatives are present, 
in particular for $\alpha=\frac{1}{9 g^2}$ we find 
\be
\label{star-new2}
e^{-1} {\cal L} =-  \frac{M_P^2}{2}  R   - \frac{1}{4} F^2({\cal V})  - \frac12  \p \varphi \, \p \varphi
-  \frac{9 g^2}{2 } \left( 1 - e^{- \sqrt{ \frac23} \varphi}  \right)^2
 - 3 g^2 e^{- 2\sqrt{ \frac23} \varphi} \, {\cal V}^m {\cal V}_m,
\ee
where ${\cal V}_m={ A}_m-3H_m$. 
This is standard supergravity coupled to  a massive vector multiplet. 
Various modifications  can be found in  
\cite{Ferrara:2014cca,Ferrara:2013eqa,Ferrara:2014rya,Farakos:2014gba}. 
We will use the dual description (\ref{star-new2}) for our discussion of the $R+R^2$ supergravity.

\subsection{Evolution}

From the Lagrangian density (\ref{star-new2}) we see that the only way to increase the 
energy density to $M_P^4$ is by giving an initially large value to the vector ${\cal V}_m$. 
In this scenario we choose the gauge 
\be
{\cal V}_0 &=&0 ,
\ee
and we take the $z$-spatial axe  parallel to the direction of the vector
\be
{\cal V}_i &=& {\cal A}_z(t) \delta_{i}^{z} . 
\ee
By giving to the vector a non-vanishing value a direction is singled out from
the other two perpendicular in the spatial space. This implies that the metric will be described by two scale factors 
\be \label{bianchi}
ds^2 = - dt^2 + a^2(t) [dx^2 + dy^2 ] +  c^2(t) dz^2 , 
\ee
hence, an anisotropy is created.
Here we have identified this direction with the $z$-axis. 
\begin{figure} 
\centering
\begin{tabular}{cc}
{(a)} \includegraphics [scale=.6, angle=0]{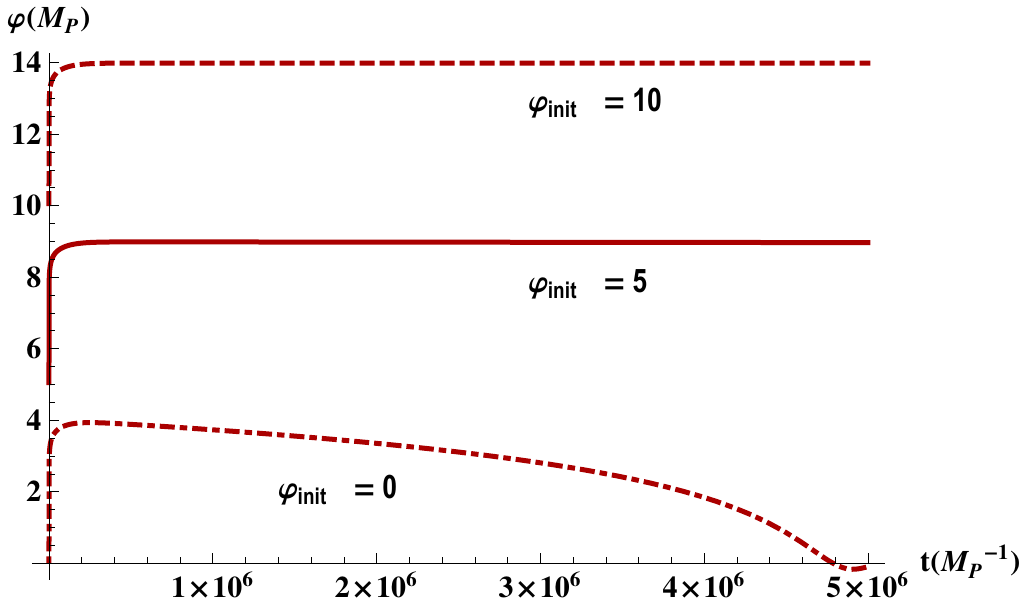} &
{(b)} \includegraphics [scale=.6, angle=0]{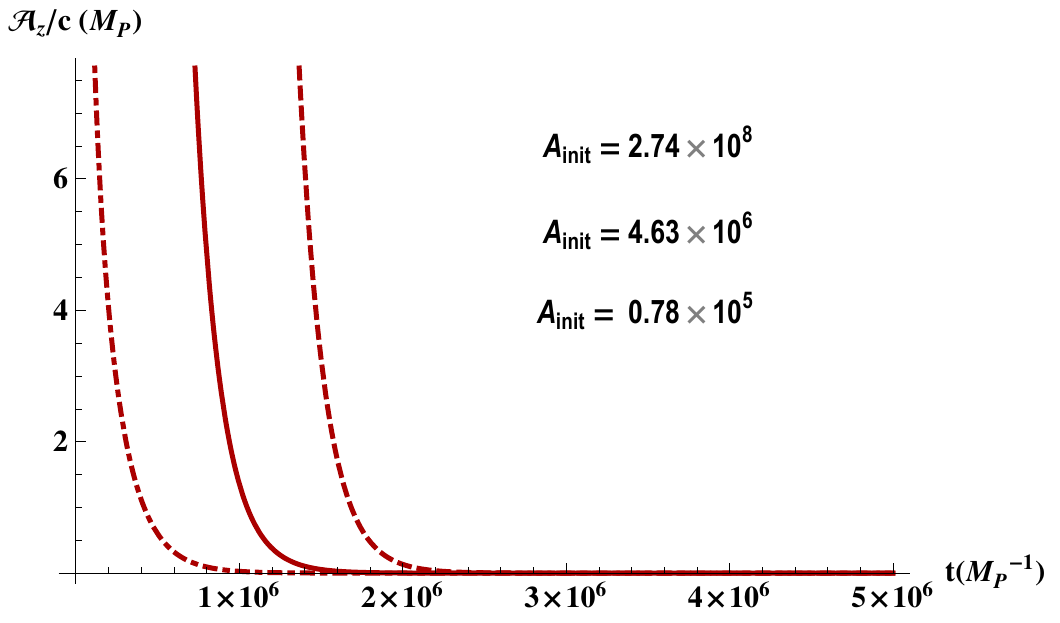}  \\
\end{tabular}
\caption{\small{The evolution of the supergravitational fields $\varphi$ (left panel) and $A_z$ (right panel) in the new-minimal embedding for different initial values and
$V(\varphi_\text{init}, A_{z,\text{init}})=\rho_\text{kin}=0.5 \rho_\text{tot,init}=\,M^4_P$. The vector field $A_z$ gets redshifted from the expansion and the scalar $\varphi$ rolls a smaller distance in field space than in the old-minimal case ruling out initial values $\varphi_\text{init}<1.5 M_P$ because not enough e-foldings are achieved. 
 Once inflation begins around $t\sim 10^5\, t_P$  the $\varphi$-field slowly rolls and the vector field $A_z$ gets completely diluted. 
}}
\end{figure}

The Einstein equations read 
\be \label{friedan1}
\left( \frac{\dot a}{a} \right)^2 + 2 \frac{\dot a}{a}  \frac{\dot c}{c}  &=&  \frac{1}{M_P^2} \rho,
\\ 
\frac{\ddot c}{c} + \frac{\ddot a}{a} +  \frac{\dot a}{a}  \frac{\dot c}{c} &=&  - \frac{1}{M_P^2} p_x = -  \frac{1}{M_P^2} p_y ,
\\ 
2 \frac{\ddot a}{a} + \left( \frac{\dot a}{a} \right)^2  &=& - \frac{1}{M_P^2} p_z,
\ee
with
\be
\rho &=&   \frac12 \dot \varphi^2 +\frac12 \left( \frac{\dot {\cal A}_z}{c} \right)^2 +V_{\text{nsugra}R^2}(\varphi, {\cal A}_z),
\\
p_x &=& p_y \,=\,\, \frac12 \dot \varphi^2 + \frac12 \left( \frac{\dot {\cal A}_z}{c} \right)^2 - V_{\text{nsugra}R^2}(\varphi, {\cal A}_z),
\\ \label{fr-an-f}
p_z &=& \frac12 \dot \varphi^2 - \frac12 \left( \frac{\dot {\cal A}_z}{c} \right)^2  - \frac92 g^2 \left( 1 - e^{-\sqrt{\frac23} \varphi/M_P} \right)^2 
+ 3 g^2 e^{- 2 \sqrt{\frac23} \varphi/M_P} \left( \frac{{\cal A}_z}{c} \right)^2.
\ee
The energy densities and pressures are defined by the energy-momentum tensor $T_{00}=\rho$, $T_{xx} =p_x/a^2$, $T_{yy}=p_y/a^2$, $T_{zz}=p_z/c^2$ for
\be
\begin{split}
T_{mn} \,=\, 
& \left(\partial_m \varphi \partial_n \varphi -\frac12 g_{mn} g^{kl}\partial_k \varphi \partial_l \varphi \right) + \left( F_{ml} F_n^{l}-\frac14 g_{mn} F_{kl}F^{kl}  \right) \\
& -g_{mn} V_{\text{nsugra}R^2}(\varphi, {\cal A}_z)+2V'_{\text{nsugra}R^2}{\cal V}_m {\cal V}_n,
\end{split}
\ee
where
\be
V_{\text{nsugra}R^2}(\varphi, {\cal A}_z)= \frac92 g^2 \left( 1 - e^{-\sqrt{\frac23} \varphi/M_P} \right)^2 
+ 3 g^2 e^{- 2 \sqrt{\frac23} \varphi/M_P} \left( \frac{{\cal A}_z}{c} \right)^2,
\ee
and the $V'_{\text{nsugra}R^2}(\varphi, {\cal A}_z)$ denotes the derivative with respect to the Lorentz invariant quantity ${\cal V}_\mu {\cal V}^\mu=({\cal A}_z/c)^2$.
Extremizing the action, the field equations read 
\be \label{scal-an}
&& \ddot \varphi +  \left(2 \frac{\dot a}{a} + \frac{\dot c}{c}  \right) \dot \varphi + \frac{\partial V_{\text{nsugra}R^2}(\varphi, {\cal A}_z)}{\partial \varphi} =0,
\\
&& \ddot {\cal A}_z +  \left(2 \frac{\dot a}{a} - \frac{\dot c}{c}  \right) \dot {\cal A}_z + \frac{\partial V_{\text{nsugra}R^2}(\varphi, {\cal A}_z)}{\partial {\cal A}_z} =0\,.
\ee
We will set
\be
g^2 = \frac16 m^2 \simeq \frac16 \, (1.3\times 10^{-5})^2\, M_P^2. 
\ee

Starting from $V(\varphi_\text{init}, {\cal A}_{z\, \text{init}}) \sim M^4_P$ the first field to roll down the slope of the potential is the scalar $\varphi$. Initially the vector field has a small mass, $m_{{\cal A}_z}\ll H$, and will stay nearly frozen. In this period the effective potential is 
\be 
V_{\text{nsugra}R^2}(\varphi) \sim V_0 e^{-2\sqrt{\frac{2}{3}}\varphi/M_P}\,.
\ee

\subsection{Initial conditions}

\begin{figure} 
\textbf{\!\!\!\!\!\!\! \!
\!\!\!\!\!\!
\!\!\!\!\!\! Evolution of the scale factor \,\,\,\,\,\,\,\,\,\,\,\,\,\,\,\,\,\,\,\,\,\,\,\,\,\, The event horizon}
\centering
\begin{tabular}{cc}
{(a)} \includegraphics [scale=.65, angle=0]{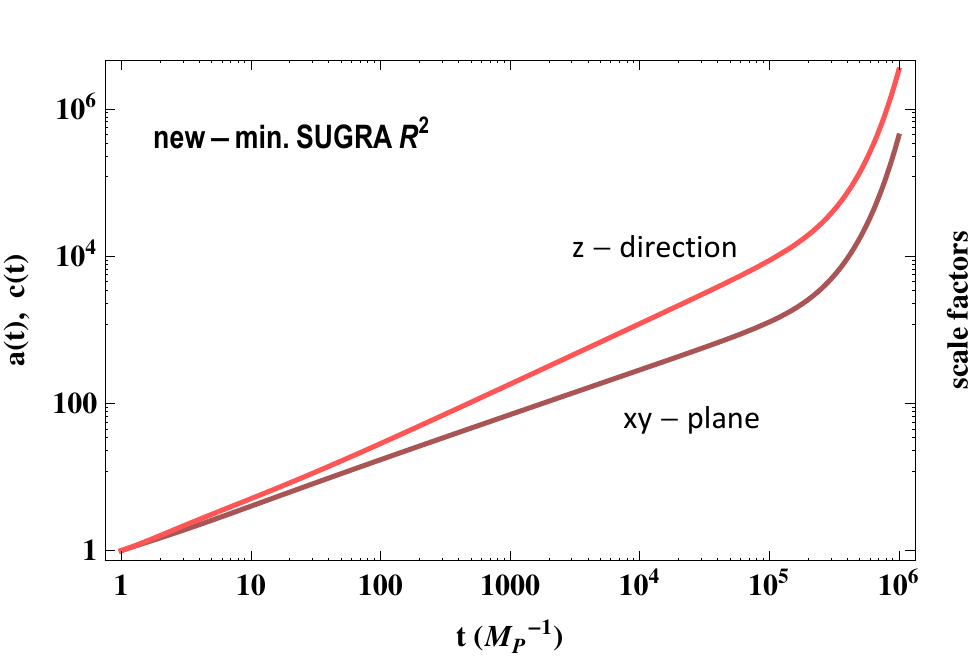} &
{(b)} \includegraphics [scale=.65, angle=0]{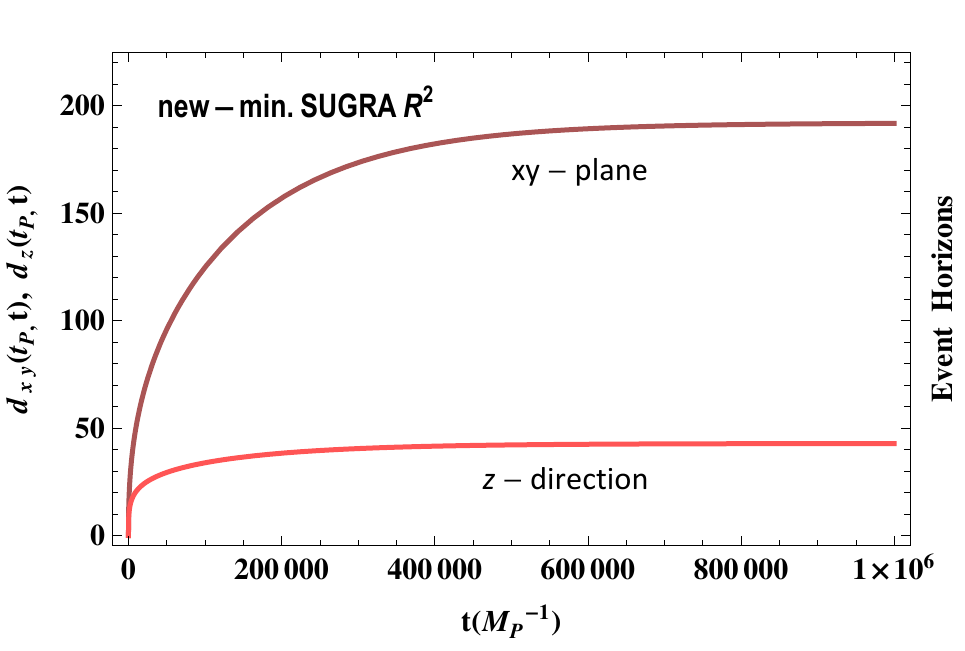}  \\
\end{tabular}
\caption{\small{The evolution of the scale factor (left panel) and the event horizon (right panel). The anisotropic expansion of the Universe is manifest which implies that the event horizon distance is respectively anisotropic. After the onset of inflation $t_\text{INF}$ the scale factors evolve similarly and the anisotropy gets diluted. The event horizon distance is in $H^{-1}(t_P)=\sqrt{3} \, l_P$ units.}}
\end{figure}

The evolution appears similar to the old-minimal case however, here, there is a background vector field with non-vanishing value which breaks the isotropy of the space and, as described in eq. (\ref{friedan1}) -(\ref{fr-an-f}), the scale factor in the directions parallel and perpendicular to the vector evolves differently. 
The expansion rate of each direction depends on the initial conditions. When $\rho_\text{init} \simeq V_\text{init} \gg \rho_\text{kin, init}$ the $z$-component of the pressure, $p_z$, is positive while the $p_x$ and $p_y$ are negative implying that the $c(t)$ scale factor grows faster than the $a(t)$ giving at the Universe a cigar-like shape. On the contrary, if $\rho_\text{init} \simeq  \rho_\text{kin, init} \gg V_\text{init}$ the $z$-component of the pressure is negative while the $p_x$ and $p_y$ are positive and the $a(t)$ scale factor grows faster than the $c(t)$ and the shape of the Universe is a pancake-like one.
We numerically solve the system of the equations and the evolution of the two scale factors, for $\rho_\text{kin, init}=(1/2) V_\text{init}$ can be seen in the Fig. 9. Accordingly, the event horizon distances change in the $z$-direction and the $x-y$ plane. The volume of the event horizon is albeit found not to be much sensitive to the partition of the initial energy density.

We can also redefine the two scale factors as $a=e^{\alpha+\beta}$ and $c=e^{\alpha-2\beta}$ where the $e^\alpha$ can be seen as the isotropic scale factor and $\beta$ the deviation from isotropy. The zero-zero component of the Einstein equation (\ref{friedan1}) can be recast into  \cite{Kolb:1990vq}
\be \label{friedan2}
{\theta}^2 =\frac{\rho}{3M^2_P}+\frac{\sigma^2}{3} \equiv \frac{1}{3M^2_P} \left(\rho+\rho_\text{AN} \right)\,.
\ee
The $3 {\theta}\equiv (2\dot a/a+\dot c/c)=3\dot \alpha$ represents the volume expansion rate, $\dot{Vol}/Vol$, and $\sigma^2=3\dot \beta^2$ is the one half of the sum of the shear components, $\sigma_i$ squared, where $\sigma_i=\dot {a}_i/a_i-\theta$. If the energy density of the vector field that feeds the anisotropy is vanishing then the background Einstein equations imply that $\dot{\beta} \propto (e^{\alpha})^{-3}$, hence the effective energy density of the anisotropy, $\rho_\text{AN} \equiv M^2_P \sigma^2$, redshifts rather fast, similar to the kinetic energy density redshift.

The scalar field equation of motion reads
\be
&& \ddot \varphi +  3\theta\dot \varphi + \frac{\partial V_{\text{nsugra}R^2}(\varphi, {\cal A}_z)}{\partial \varphi} =0\,.
\ee
As we can see from the analogue of the Friedman equation (\ref{friedan2}) the Hubble friction has been replaced by the "volume expansion friction"  which is larger than the usual value $(\rho/3M_P)^{1/2}$. The anisotropy aids the slow roll of the $\varphi$ field, which at first sight may look somewhat helpful for the amelioration of the initial conditions problem. 
Although it would be rather interesting to find a slow roll regime due to the presence of the $\rho_\text{AN}$, the anisotropy does not yield any helpful increase in the expansion rate of the Universe. The numerical results are shown in Fig. 8 and 9.
We finally mention that the primordial anisotropy described by the metric (\ref{bianchi}) corresponds to a Bianchi spacetime which does not prevent inflation from starting \cite{Goldwirth:1991rj}, a result which is also manifest in Fig 9.

The minimum homogeneous region required at $t_\text{init}=t_P$ for inflation to start at $t_\text{INF}$ in the new-minimal supergravity has volume
\be \label{oblate}
\begin{split}
V_\text{homog}(t_P) \, =\, 
& \frac43 \pi\, \left(d^{\,xy}_\text{event} (t_P, t_\text{max}) +  \, H^{-1}(t_\text{INF})\frac{a(t_P)}{a(t_{INF})} \right)^2  \\
& \times \, \left(d^{\,z}_\text{event} (t_P, t_\text{max})  + \, H^{-1}(t_\text{INF})\frac{c(t_P)}{c(t_{INF})}  \right) 
\end{split}
\ee
where $d^{xy}_\text{event}$ the event horizon in the  $x-y$ plane and $d^{z}_\text{event}$ in the $z$-direction. At the onset of inflation the contribution of the vector fields in the energy density is subdominant and the initial patch that gets inflated has a spherical volume $V=\frac43 \pi H^{-3}(t_\text{INF})$. However, the initial homogeneous volume (\ref{oblate}) is an oblate spheroid.

The number of the causally disconnected regions is found numerically to be of the order 
\be
\frac{V_\text{homog}(t_P)}{\frac43 \pi l_P^3} \, \sim \, 10^7 \,\, \text{CDR}\,.
\ee

\begin{table} \label{tabb2}
{\small
\begin{center}
$$
\begin{array}{|r|r|r|r|r|}\hline
 \begin{split}
 & \quad  \bold{Initial\,Values \,(old)}\\
 &  V_{\text{sugra}R^2}(\vphi_\text{init}, b_\text{init})\sim 1 
 \end{split} 
 \quad &  
 \begin{split}
 & \quad  \bold{Initial\,Values\, (new)}\\
 &  V_{\text{nsugra}R^2}(\vphi_\text{init}, A_\text{init})\sim 1 
 \end{split} 
 \quad & 
 N \gtrsim 60 \, & \bold{Selfrepr.} & \varphi_\text{INF} \quad\quad\quad\quad \\
\hline 
\hline
  \begin{split} 
  &\quad \varphi_\text{init} \gtrsim   4 \quad\quad \\
  &  b_\text{init} \gtrsim \, 2 \times 10^6 \quad\quad
  \end{split}  \quad &  
   \begin{split} 
  &\varphi_\text{init} \gtrsim   13.5 \quad\quad  \\
  &  A_\text{init} \gtrsim \, 5\times 10^9 \quad\quad 
  \end{split}  \quad &
    \text{Yes} \quad \,  & \text{Yes}\quad  & \varphi_{INF} \gtrsim 17.5\quad \quad \\
\hline
\begin{split} 
  & -8.5 \lesssim \varphi_\text{init} \lesssim   4 \quad\\
  &     10^2   \lesssim b_\text{init} \lesssim 2\times 10^6  \quad
  \end{split} \quad  &
  \begin{split} 
  & \quad \quad 1.5 \lesssim \varphi_\text{init} \lesssim   13.5 \quad \\
  &  3\times 10^5 \lesssim  A_\text{init} \lesssim 5\times 10^9  \quad
  \end{split} \quad  &
   \text{Yes} \quad \,   & \text{No} \quad  &  5.5 \lesssim \varphi_{INF} \lesssim 17.5  \\
\hline
\begin{split} 
  & \varphi_\text{init} \lesssim   -8.5 \quad\quad \\
  &   b_\text{init} \gtrsim \, 10^2 \quad\quad
  \end{split}  \quad  &
  \begin{split} 
  & \varphi_\text{init} \lesssim   1.5 \quad\quad \\
  &  A_\text{init} \gtrsim \, 3\times 10^5 \quad\quad 
  \end{split}  \quad  &
  \text{No}\quad \, & \text{No}\quad  & \varphi_{INF} \lesssim  5.5 \quad\quad  \\
\hline
\end{array}
$$
\end{center}
\caption{\small The table presents the constraints on the (old- and new-minimal) supergravity scalar field space, in units $M_P=1$, for sufficient number of e-foldings starting from Planck-scale initial energy densities for the potential and the kinetic terms. The field values that lead to a selfreproduction regime are also shown. } }
\end{table}

\section{The curvature term}
An homogeneous and isotropic spacetime is described by the FLRW metric
\begin{equation}
ds^2=-dt^2+a^2(t)\left(\frac{dr^2}{1-kr^2}+r^2(d\theta^2+sin^2\theta d\phi^2) \right),
\end{equation}
and the evolution of the scale factor is given by the Friedmann equation
\begin{equation} \label{F}
\left(\frac{\dot{a}}{a}\right)^2 =\frac{\rho}{3M^2_P} - \frac{k}{a^2}.
\end{equation}
When $k \neq 0$ the space is curved and the volume enclosed in sphere of radius $d=a\, \chi_0$ differs from that of the flat space. It is \cite{Mukhanov:2005sc}
\be \label{volu}
V= 
\begin{cases} 
\,V_\text{closed}\,=\,\pi a^3(2 \chi_0-\sin 2\chi_0)\,, \quad\quad\, 0\leq\chi_0\leq \pi   \\ 
\,V_\text{flat}\,=\,\frac43\pi a^3\,,  \quad\quad\quad\quad\quad\quad\quad\quad\quad\,\,   \chi_0=d_\text{com}  \\
\,V_\text{open}\,=\,\pi a^3(\sinh 2\chi_0-2\chi_0)\,, \quad\quad            0\leq\chi_0\leq \infty\,.    \\
\end{cases}
\ee
In the closed and the open case the  $\chi_0$ is a dimensionless anglular coordinate and the scale factor $a$ is the dimensionful radial coordinate. In the flat space $\chi_0$ corresponds to a dimensionful comoving distance. For a definite $\chi_0$ it is
\be \label{Vh}
V_\text{closed} (\chi_0) < V_\text{flat} (\chi_0) < V_\text{open} (\chi_0)\,.
\ee
When $\chi_0 \ll 1$ the volumes do not differ much while for $\chi_0 > 1$ the difference becomes important.

If the energy density of the Universe is larger than the critical density, $\rho_\text{crit}=3H^2M^2_P$,  the Universe has a closed spatial geometry ($k=1$), whereas for $\rho<\rho_\text{crit}$  an open ($k=-1$) one. In the special case that $\rho=\rho_\text{crit}$ we have a flat geometry ($k=0$). In the preceding sections we have implicitly assumed that in the pre-inflationary stage it is $|\rho-\rho_\text{crit}| \ll 1$. This accounts for a rather special initial condition.  Nevertheless the results obtained can be accordingly translated into the case that the spatial curvature is nonzero. We mention that the present data find no evidence for any departure from a spatially flat geometry \cite{Planck:2015xua}
\be
|\Omega_K|<0.005\,,
\ee
where $\Omega_K=1-\rho_\text{tot}/\rho_\text{crit}$. It is actually inflation itself that addresses the puzzle of the observed flatness of the Universe. The $\Omega_K$ decreases exponentially with time during the accelerated expansion. In the models studied here it is very natural to expect many more than $60$ e-foldings and this fact diminishes the chances for any observable deviation from flatness. However, before inflation a homogeneous initial patch is expected to feature either a closed or an open FLRW geometry.

\subsection{Closed Universe}
When the Universe has a positively curved geometry, the Friedmann equation can be written as
\begin{equation}
3M^2_PH^2=\rho-\rho_\text{closed},
\end{equation}
where $\rho_\text{closed}\equiv 3M^2_P/a^2$. At the moment $t_\text{turn}$ the Universe reaches its maximum size and $\dot{a}(t_\text{turn})=0$. There, the energy density has the value $\rho(t_\text{turn})=3M^2_P/a^2_\text{turn}$ and the evolution turns from  expansion to collapse. Inflation has to start before $t_\text{turn}$, that is $\rho(t_\text{INF})>\rho_\text{closed}$. Assuming a scaling of the energy density of the form 
\be \label{rho-scal}
\rho=\rho_\text{init}\left(\frac{a_\text{init}}{a} \right)^n \simeq \frac{12}{n^2}\left(\frac{t}{M_P}\right)^{-2}\,,
\ee
where the second equality follows from eq. (\ref{scalef}) for $t\gg t_\text{init}$, we find the expression for the initial radius in terms of the $a_\text{turn}$ radius 
\be \label{a-min}
a_\text{init} \, = \, \sqrt[n]{3\, M^2_P} \,a_\text{turn}^{1-2/n}\,.
\ee
The turnover is postponed to the far future if $\rho(t_\text{INF})\gtrsim 3/a^2_\text{turn}$, i.e. $a_\text{turn} \gtrsim (n/2)\, t_\text{INF}$, that gives the relation between the time of inflation and the minimum initial radius, $a(t=t_\text{init})\equiv a_\text{init}$, of the closed Universe
\be \label{a-min-t}
a_\text{init,min} \sim  \sqrt[n]{3\, M^2_P} \left(\frac{n}{2}\right)^{1-2/n}\,t_\text{INF}^{1-2/n}\,.
\ee
The above expression (\ref{a-min-t}) is an approximate one because a subsequent accelerating phase takes over. The equation of state changes continuously from the value $w=n/3-1$ to the value $w\leq -1/3$ and the expression (\ref{a-min}) is not exact when the scaling of the energy density deviates from (\ref{rho-scal}).  The (\ref{a-min-t}) yields an overestimated value.  Below we present the cases with and without supergravity separately.

\subsubsection{$R+R^2$}
When the inflationary energy density is bounded from above, as it is in the Starobinsky model, $V_\text{INF} \lesssim 1.2 \times 10^{-10} M^4_P$, then apparently, for our Universe to survive,   the turnover energy density has to be in lower values: $\rho_\text{turn} < V_\text{INF} \simeq 1.2 \times 10^{-10}M^4_P$. An approximate estimation is $a_\text{turn} > {\cal O} (\sqrt{3M^2_P/V_\text{INF}})$. The numerical estimation yields  $a_\text{turn} >\,a(t_\text{INF}) \simeq 1.3 \times 10^5\, l_P $ and $t_\text{INF} \simeq 0.6 \times\, 10^{5} \,t_P$. 
Before inflation, higher energy densities are in the form of kinetic energy of the scalaron field, hence
\begin{equation}
\rho=\left(\frac12 \dot{\varphi}^2_\text{init} \left(\frac{a_\text{init}}{a(t)} \right)^6 +V_{R^2}(\varphi) \right)\simeq \rho_\text{kin,init}\left(\frac{a_\text{init}}{a(t)} \right)^6.
\end{equation}
For $\rho_\text{kin,init}=M^4_P$ and from (\ref{a-min}) we take an estimation for the value of the initial radius of the closed Universe $a_\text{init} \sim \,a^{2/3}_\text{turn}\, l_P^{1/3}$.  
 Numerically we find 
\begin{equation} \label{cl_rad} 
a_\text{init} \geq a_\text{init,min}\sim 2.5 \times 10^{3}\, l_P\,.																																	 
\end{equation}
This is a rather large radius. It means that when the initial Universe emerged from the Planck, possibly quantum gravity, era it must have had a radius $a(t_P)= a_\text{init}$ of at least a few thousand times the fundamental Planck length. This is a formidable radius for theories that attempt a description of the quantum genesis of our Universe \cite{Linde:1983cm, Vilenkin:1984wp, Linde:2005ht}. If it had any smaller size it would have collapsed before inflation begins.

The minimal size of the initial radius $a_\text{init}$ depends on the time inflation starts. For a Universe dominated by the kinetic energy, the energy density falls like $\rho \propto a^{-6}$ and from (\ref{a-min-t}) we take for $t_\text{INF}\gg t_\text{init}=t_P$
\be
a_\text{init,min}(t_\text{init} =t_P, t_\text{INF})  \sim \, t^{2/3}_\text{INF} \,\,l^{1/3}_P.
\ee

We have implicitly assumed that the homogeneous volume is the entire volume of the closed Universe, $2\pi^2 a_\text{init}^3$, because we considered an FLRW evolution for the whole spacetime not only for an initial patch.  Here as well, the closed Universe with radius (\ref{cl_rad}) contains billions of initially causally disconnected regions. The minimum volume (\ref{volu}) that can be considered to be in causal contact is the one enclosed in a sphere of radius  $d=a_\text{init,min} \chi_0 =l_P$. It is $\chi_0=l_P/a_\text{init,min}\sim 1/2500 \ll 1$ hence 
\be
V_\text{closed}(a_\text{init}\, \chi_0=l_P) \simeq \frac43 \pi l^3_P \, \ll \, V_\text{closed}(a_\text{init}\, \pi) = 2\pi^2 a_\text{init}^3\,.
\ee
The entire volume $V^\text{tot}_\text{closed}= V_\text{closed}(a_\text{init,min}\, \pi)$ 
contains
\be \label{cCDR}
\frac{V^\text{tot}_\text{closed}}{V_\text{closed}(l_P)} \simeq\frac{ 2\pi^2 a^3_\text{init,min}}{\frac43 \pi l^3_P }\sim 7\times 10^{10}\,\, \text{CDR}\,, 
\ee
homogeneous causally disconnected regions (CDR) for $t_\text{init}=t_P$ as  depicted in Fig. 10.

\subsubsection{$R+R^2$ supergravity}
In the Starobinsky supergravity model, potential energy density values $V_{\text{sugra} R^2}(\vphi) \gg V_\text{INF}$ are possible\footnote{In the supergravity case also holds $V_\text{INF} \ll M^4_P$, however $V_{\text{sugra} R^2}(\vphi) \gg V_\text{INF}$ values do not rule out inflation as it happens in the conventional Starobinsky model.}. We assume that the potential energy after the Planck era is of the order of the Planck mass to the fourth power, $\rho_\text{init}= M_P^4$. An approximation for the scaling of the supergravity energy density is given by the expressions (\ref{approxsugra}) that is
\begin{equation}
\rho_\text{sugra}\sim \rho_{w=-1/9}= \rho_\text{init} \left(\frac{a_\text{init}}{a}\right)^{8/3}.
\end{equation}
Hence, $a_\text{init} \sim \,\,a^{1/4}_\text{turn}\,\, l^{3/4}_P$.  Asking again for $\rho_\text{turn} < V_\text{INF} \simeq 1.2  \times 10^{-10} M_P^4$  we find an estimation for the minimum initial radius.
 The dependence of the $a_\text{init,min}$ on the time inflation starts 
   has the following form for $t_\text{INF}\gg t_\text{init}=t_P$
\be
a_\text{init,min} (t_\text{init}=t_P, t_\text{INF}) \sim  \,  t^{1/4}_\text{INF} \,l^{3/4}_P \, , \quad\quad \text{for} \quad w=-1/9\,.
\ee
The above accounts for a conservative approximation.
Numerical estimations for the supergravitational system yield $a_\text{turn} > \, a(t_\text{INF}) \simeq 10^5 \, l_P$, $t_\text{INF}\simeq 0.74\times 10^{5}\, t_P$ and the lower bound on the minimum radius of the 3-sphere is 
\begin{equation} \label {radsugra}
a_\text{init} \geq a_\text{init,min}  \simeq \,1.7 \times 10  \,l_P\, , \quad\quad \text{for} \quad w_{\text{sugra}\,R^2}\,.
\end{equation}
We mention that  when the spatial geometry is curved the scale factor corresponds to the radial coordinate and it is dimensionful. The ratio $a(t)/a_\text{init}$ denotes the number of times the radius of the closed Universe has increased compared to the initial radius.

Compared to (\ref{cl_rad}) the bound (\ref{radsugra}) on $a_\text{init}=a(t_P)$  accounts for more than one hundred times less severe condition for a closed Universe to reach the inflationary period before it starts collapsing. Again here, we have assumed that the entire space is an homogeneous three-dimensional sphere with thousands of initially disconnected regions. 
The minimum volume (\ref{volu}) that can be in causal contact is enclosed in a sphere of radius $a_\text{init,min} \chi_0 =l_P$. It is $\chi_0=l_P/a_\text{init,min} \sim 1/17$ and also here
\be
V_\text{closed}(a_\text{init}\, \chi_0=l_P) \simeq \frac 43 \pi l^3_P \, \ll \, V_\text{closed}(a_\text{init}\, \pi) = 2\pi^2 a_\text{init}^3\,.
\ee
The entire volume $V^\text{tot}_\text{closed}= V_\text{closed}(a_\text{init,min}\, \pi)$ 
contains  
\be
\frac{V^\text{tot}_\text{closed}}{V_\text{closed}(l_P)} \simeq \frac{ 2\pi^2 a^3_\text{init,min}}{\frac43 \pi l^3_P }\sim 2\times 10^4\,\, \text{CDR} \,,
\ee
for $t_\text{init}=t_P$, as  depicted in Fig. 10, that is few million times less CDR than the $R^2$ case.

\subsection{Open Universe}

The initial patch that has been inflated may locally resemble a geometry of negative curvature. In this case the Friedmann equation reads
\begin{equation}
3M^2_PH^2=\rho+\rho_\text{open}\,,
\end{equation}
where $\rho_\text{open}\equiv 3M^2_P/a^2$.
Here, the space corresponds to a hyperbolic plane that has an infinite volume, though we are interested only in the local geometry of space inside the Hubble radius not globally. The scale factor can take small values without any fear that the Universe will collapse. The $\rho_\text{open}$ can dominate over the energy density of the Universe and dilute the matter till the energy is redshifted to the value of the inflationary plateau $V_\text{INF}$. Then the inflationary evolution takes over, spacetime becomes approximately de Sitter and the negative curvature term, represented by the $\rho_\text{open}$, asymptotically vanishes.

This case yields a rather small event horizon distance. In particular the curvature term scales like $1/a^2$ and the $a(t)$ grows like $t$. In the eq. (\ref{devent}) the power is $n=1$ and the event horizon untill inflation grows only logarithmically with time
\begin{equation}
d_\text{event}(t_\text{init},\, t_\text{max} )= a(t_\text{init})\int_{t_\text{init}}^{t_\text{max}} \frac{dt}{a(t)} =a(t_\text{init}) \left[\, \ln \left(\frac{t_\text{INF}}{t_\text{init}} \right)+ \int_{t_\text{INF}}^{t_\text{max}}\frac{dt}{a(t)} \,\right]\,.
\end{equation}
For $t_\text{max} \gg 10^5 t_P\sim t_\text{INF}$ we find numerically
\begin{equation} 
d_\text{event}(t_\text{init}=t_P, \, t_\text{max}>t_\text{INF} ) \sim 12.5  \, a(t_P)\, ,
\end{equation}
where $a(t_\text{init}=t_P)=a(t_P)$ is the real value of the hyperbolic space radius. The minimum required homogeneous regions has radius
\be \label{d-open}
D_\text{homog}(t_P) = d_\text{event} (t_P, t_\text{max}) +  \, H^{-1}(t_\text{INF})\frac{a(t_\text{P})}{a(t_{INF})} \sim 14 \,a(t_P)\,,
\ee
since $a(t)\sim t$. 
The (\ref{d-open}) seems to be a comfortably small value. The imaginary radius of the open Universe can take small values such as the Planck length and expand fast and endlessly resulting, at first sight, in only a ``logarithmic sensitivity" of the initial patch to the regions that can influence it. 
However, if the initial radius of curvature $a(t_P)$ is about the Planck length then the volume enclosed in a sphere of radius $D_\text{homog}\sim 14\, a(t_P)$ is remarkably large in such a highly curved hyperbolic space. 
Indeed it has to be $a(t_P)\chi_0 \sim 14\, a(t_P)$ hence  the angular coordinate is $\chi_0\sim 14 \gg 1$ and the Euclidean space approximation breaks down.
It is
\be
V_\text{open} \left( a(t_P)\chi_0=14 a(t_P) \right)\,=\,\pi a^3(t_P)(\sinh 2\chi_0-2\chi_0)|_{\chi_0 =14} \, \simeq   \, (7.2 \times 10^{11})\,\pi a^3(t_P)\, .
\ee
On the other hand, the minimum volume of a causally connected region is enclosed in a radius about $l_P$ and has size
\be
V_\text{open}\left(a(t_P)\chi_0\sim l_P\right)\,=\,\pi a^3(t_P)(\sinh 2\chi_0-2\chi_0)|_{\chi_0 =\frac{1}{\sqrt{3}}}\, \simeq  \, 0.27 \,\pi a^3(t_P) \, .
\ee
We took $a_\text{init}=a(t_P)=l_P\,\sqrt{3}$ in order that $H^2(t_P)=1/(3 l_P^2)$ and be in agreement with the flat and close Universe cases examined before where $H^2(t_P)\simeq M^2_P/3$.
Hence, we find 
\be \label{opCDR}
\frac{V_\text{open}(D_\text{homog})}{V_\text{open}(l_P)} \sim 2.7 \times 10^{12} \,\,\text{CDR}\,, 
\ee
at $t_\text{init}=t_P$. This is a very large number. 
The (\ref{opCDR}) is actually larger than the CDR numbers the Starobinsky $R^2$ model yielded for the cases of flat (\ref{fCDR}) or closed (\ref{cCDR}) Universe. We note that in the closed Universe case the initial radius had to be large enough to evade collapse and, hence, the curvature term was subdominant.  Here the space is hyperbolic and the Euclidean approximation cannot be applied for distances larger than the Planck length.  

If we consider a subdominant curvature term, $\rho_\text{open}=3M^2_P/a^2 \ll \rho$ then the flat space approximation is reliable and the results should be similar to those of flat and large radius closed Universe.
In particular, for the $R+R^2$ (super)gravity model, the $D_\text{homog}$ should not be much different for $k=0, \pm1$. 
Then according to the hierarchy of the volumes in these three different geometries (\ref{Vh})  we take again that
\be \label{hier}
\# \,\text{CDR}(\text{closed}) < \# \,\text{CDR} (\text{flat}) < \# \,\text{CDR}(\text{open})\,.
\ee
Hence the hierarchy (\ref{hier}) holds either for a dominant or subdominant  $k=-1$ curvature term.

\section{Conclusions}

In this work we studied the $R+R^2$ gravity and supergravity models for inflation (known also as Starobinsky models),  which are  characterized by a plateau inflationary potential and are particularly motivated after the release of the P{\scshape lanck} 2013 results. However,  they account for low energy scale  inflaton models, requiring a rather extended acausal homogeneity in order for inflation to occur. We demonstrated that the problematic issue  of the initial conditions is less severe if supergravity is realized in nature due to the extra directions in the field space that can implement a relatively fast expansion rate before inflation. For flat  (closed) background geometry for the Universe, the $R+R^2$ gravity requires a huge initial homogeneous patch (huge initial 3-sphere) that contains about $10^{11}$ causally disconnected sub-patches while in the $R+R^2$ supergravity this number is  at least $10^6$ times smaller. 

We considered topologically trivial FLRW geometries.  The homogeneous patch of radius $D_\text{homog}$  is enclosed in a smaller volume when $k=1$ and in a larger one when $k=-1$. The level of fine tuning is the minimum one when the background spatial geometry has a positive curvature. Hence the $k=1$ case seems to be favoured unless the Universe is described by a non-trivial topology.

Also, we mention that the study of the pre-inflation supergravitational dynamics revealed interesting features such as the initial conditions that give sufficient number of e-foldings, that can  avoid the eternal process of self-reproduction, and generate a remarkable, however ephemeral, anisotropy. 

We underline that throughout this work we focused on pure $R+R^2$ gravitational and supergravitational settings assuming that the dynamics of the higher curvature terms are solely responsible for the early Universe inflationary phase. 
Under these assumptions, our results, regarding the initial conditions problem and for a trivial topology, point towards $k=1$ and theories with extra field dimensions such as the $R+R^2$ supergravity theory.

\begin{figure} 
\textbf{\,\,\,\,\,\,\,\,\,\,\,\,\,\,\,\,\,\,\, Number of the causally disconnected regions (\# CDR) \,\,\,\,\;\;\;\;\;\;\;\quad \quad}
\centering
\includegraphics [scale=.9, angle=0]{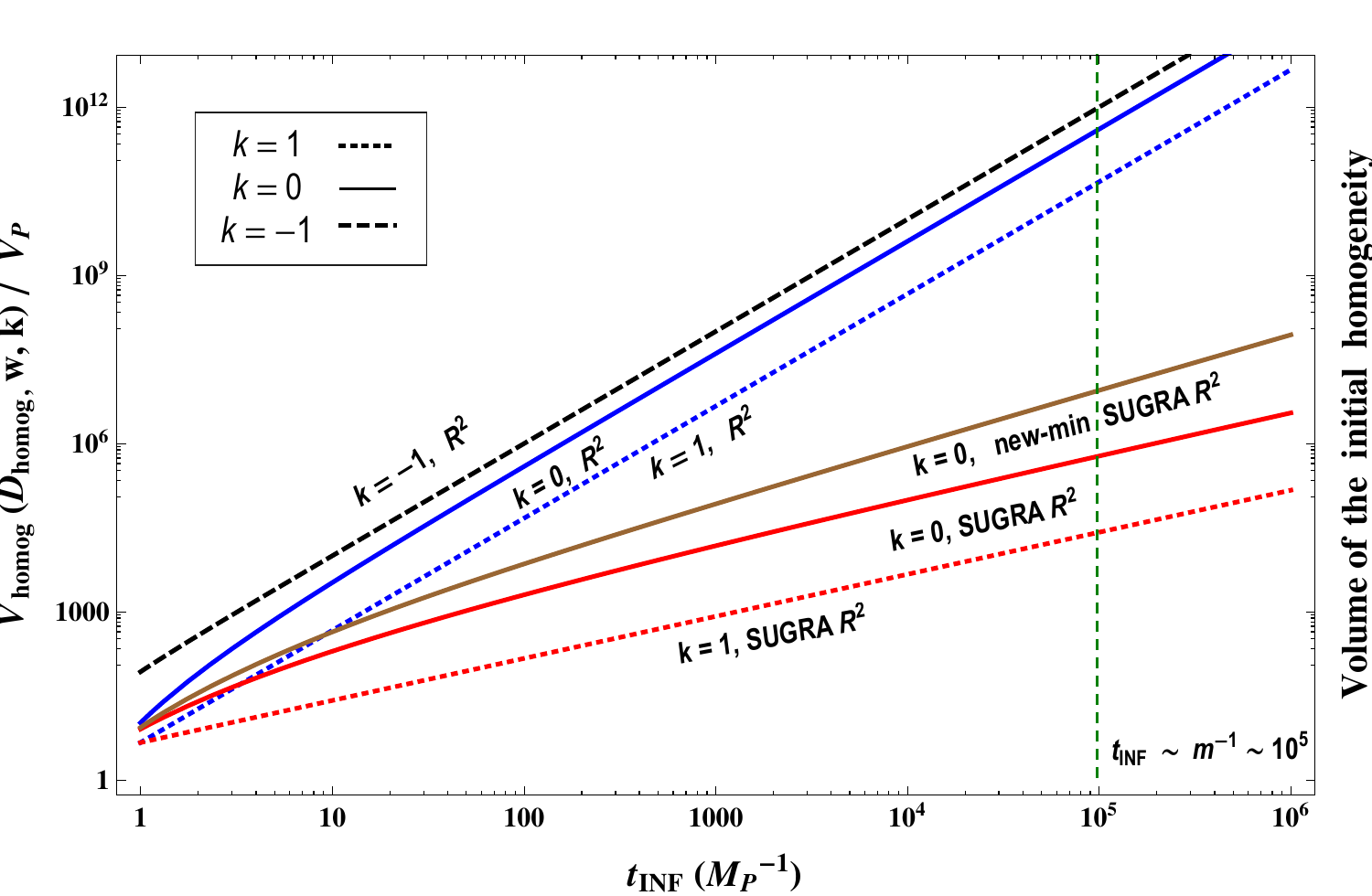} 
\caption {\small {The figure depicts the number of the causally disconnected regions (\# CDR) right after the Planck time required to be homogeneous in order for inflation to start. 
The horizontal axis is the time that inflation starts (for the $R+R^2$ (super)gravity models the $t_\text{INF}$ is fixed by the CMB with $t_\text{INF}\sim m^{-1}\sim 10^5\, t_P$; the plot manifests the initial conditions problem for the low-scale inflationary models generally).
The solid lines correspond to a flat background spatial geometry, the dotted to a positively-curved and the dashed to a negatively-curved one. The red lines correspond to $R+R^2$ old-minimal suprgravity, the brown to new-minimal supergravity and the blue to $R+R^2$ gravity. It is shown that the extra supergravity fields decrease the number of CDR.  For the $k=-1$ case the curvature dominates over the fields until the onset of inflation. At  $t_\text{INF}=t_P$ the CDR number is nonzero; the different initial CDR numbers at $t_\text{INF}=t_P$ is solely due to the background curvature. The CDR number is always larger for the $k=-1$ and smaller for the $k=1$ background geometry regardless the time inflation occurs. 
}}
\end{figure}

\section*{Acknowledgments}
\vspace*{.5cm}
\noindent 
We thank Alex Kehagias for discussion. 
I.D. would like to thank CERN theory division for hospitality during the preparation of this work.
The work of F.F. is supported by the Grant agency of the Czech republic under the grant P201/12/G028.

\end{document}